\documentclass{emulateapj}
\usepackage{natbib,times}
\usepackage{lscape}

\shorttitle{An efficient method to identify galaxy clusters}
\shortauthors{W. W. Xu et al.}
\begin{document}

\title{An efficient method to identify galaxy clusters by using
  SuperCOSMOS, 2MASS and WISE data}

\author{W. W. Xu\altaffilmark{1,2},
        Z. L. Wen\altaffilmark{1}
        and 
        J. L. Han\altaffilmark{1}}

\altaffiltext{1}{National Astronomical Observatories, 
                 Chinese Academy of Sciences, Beijing 100012, China; 
                 zhonglue@nao.cas.cn.}
\altaffiltext{2}{University of Chinese Academy of Sciences, Beijing 100012, China}


\begin{abstract} 

The survey data of Wide-field Infrared Survey Explorer (WISE) provide
an opportunity for the identification of galaxy clusters. We present an
efficient method for detecting galaxy clusters by combining the WISE
data with SuperCOSMOS and 2MASS data. After performing star-galaxy
separation, we calculate the number of companion galaxies around the 
galaxies with photometric redshifts previously estimated by the
SuperCOSMOS, 2MASS and WISE data. A scaled richness is then defined to
identify clusters. From a sky area of 275 ${\deg}^2$ coincident with
Sloan Digital Sky Survey Stripe 82 region, we identify 302 clusters in
the redshift range of $0.1<z<0.35$, 247 (82$\%$) of which are
previously known SDSS clusters. The results indicate that our method
is efficient for identifying galaxy clusters by using the all sky data of
the SuperCOSMOS, 2MASS and WISE.

\end{abstract}


\section{Introduction}

Galaxy clusters are known as the largest gravitational bounded systems
in the universe. They are located at nods of cosmic web. The space
distribution of galaxy clusters traces the large scale structure
\citep{bah88,aem11,hhw+12}. Their mass distribution can be used to
constrain cosmological parameters \citep{rb02,whl10}. Clusters are
also regarded as natural gravitational lens to magnify faint
background sources \citep{whj11,lwh+13} and laboratories to study
galaxy evolution \citep{wh11,lwh+12}. Discovery of galaxy clusters is
the basis for many related studies.

In optical/infrared, a lot of methods have been used to detect
clusters from image data. By visual inspection of optical images,
Abell identified more than 4000 nearby rich clusters covering the
whole sky \citep{abe58,aco89}. Similar visual inspections were
undertaken by \citet{zhw68} and \citet{gho86}. For reducing
subjectivity, automated peak-finding methods were developed, e.g.,
matched-filter algorithm \citep{plg+96}, adaptive kernel technique
\citep{gdl+03} and Voronoi tessellation techniques
\citep{rbf+01,kkp+02}. The single-band image data always suffer severe
contamination from foreground and background galaxies and provide very
poor redshift estimation of clusters.

The Sloan Digital Sky Survey \citep[SDSS,][]{yaa+00} offers an
opportunity to identify a large number of clusters. It provides
photometry in five broad bands ($u$, $g$, $r$, $i$, and $z$) covering
$\sim$ 14,000 deg$^2$, as well as the follow-up spectroscopic
observations. In such multi-band surveys, galaxy colors are related to
their redshifts. Member galaxies have similar colors and show a red
sequence, so that proper cuts in colors can reduce projection effect
for cluster identification. A lot of red-sequence based methods have
been developed \citep{gy00,gy05,gsn+02,kma+07}. Photometric redshifts
were estimated for all galaxies \citep{cbc+03,olc+08}. A large number
of galaxy clusters have been found based on photometric redshift
\citep{whl09,whl12,spd+11}.

Except for Abell clusters at low redshift \citep{aco89}, there
is no all sky cluster catalog up to intermediate redshift of
$z\sim0.3$. The Wide-field Infrared Survey Explorer (WISE) is an
all-sky survey at infrared wavelengthes with the observation depth
similar to that of the SDSS \citep{ydt+13}, which provides
an opportunity to identify many new clusters covering the whole
sky. Combined with optical data, the WISE data have been used to
identify clusters at high redshift \citep{ggs+12}.


In this paper, we present a simple but very efficient method to
identify galaxy clusters by combing the WISE data with the SuperCOSMOS
\citep{him01} and Two Micron All Sky Survey \citep[2MASS,][]{scs+06}
data. In Section 2, we describe the data, present our method for the
identification of galaxy clusters, and apply it to the data in the
SDSS Stripe 82 region\footnote {http://cas.sdss.org/stripe82/en/}.  In
Section 3, we compare the identified clusters with previous SDSS
clusters. A summary is given in Section 4.

\section{Data and algorithm for cluster identification}

2MASS is an all-sky survey in three infrared bands, $J$ (1.25 $\mu$m),
$H$ (1.65 $\mu$m) and $K_s$ (2.17 $\mu$m). The effective resolution of
2MASS is $\sim$5$''$. The magnitude limits of 2MASS are 15.8, 15.1 and
14.3 (10$\sigma$) for point sources, and 15.0, 14.3 and 13.5 for
extended sources in the three bands, respectively \citep{scs+06}.

The WISE observes the whole sky in four infrared bands, $W1$ (3.4
$\mu$m), $W2$ (4.6 $\mu$m), $W3$ (12 $\mu$m) and $W4$ (22 $\mu$m). The
angular resolutions in the four bands are 6.1$''$, 6.4$''$, 6.5$''$
and 12.0$''$, respectively. The all sky survey depths are 16.5, 15.5,
11.2 and 7.9 $\rm{mag}$ (5$\sigma$) for point sources
\citep{wem+10}. The most important data for cluster identification are
WISE $W1$-band and 2MASS $J$-band magnitudes. In terms of detecting
distant galaxies, the WISE data is much deeper than the 2MASS
data. The limit of cluster detection mainly depends on the depth of
the 2MASS data.

Because of the poor resolutions, both WISE and 2MASS data have no
reliable star-galaxy separation for most objects by morphological
parameters. However, stars and galaxies can be well separated by color
index, $W1-J$ \citep{ks13}. We cross-match the 2MASS-WISE objects with
known stars and galaxies in SDSS DR7 \citep{dr7+09} brighter than
$r=21$ mag with a matching radius of 3$''$. In Figure~2, we show the
efficiency for the star-galaxy separation with the $W1-J$ color index
of the WISE-2MASS data. Galaxies are very well separated with the
criteria of $W1-J<-1.5$ mag.

\vspace*{8mm}
\begin{center}
\includegraphics[angle=270,width=70mm]{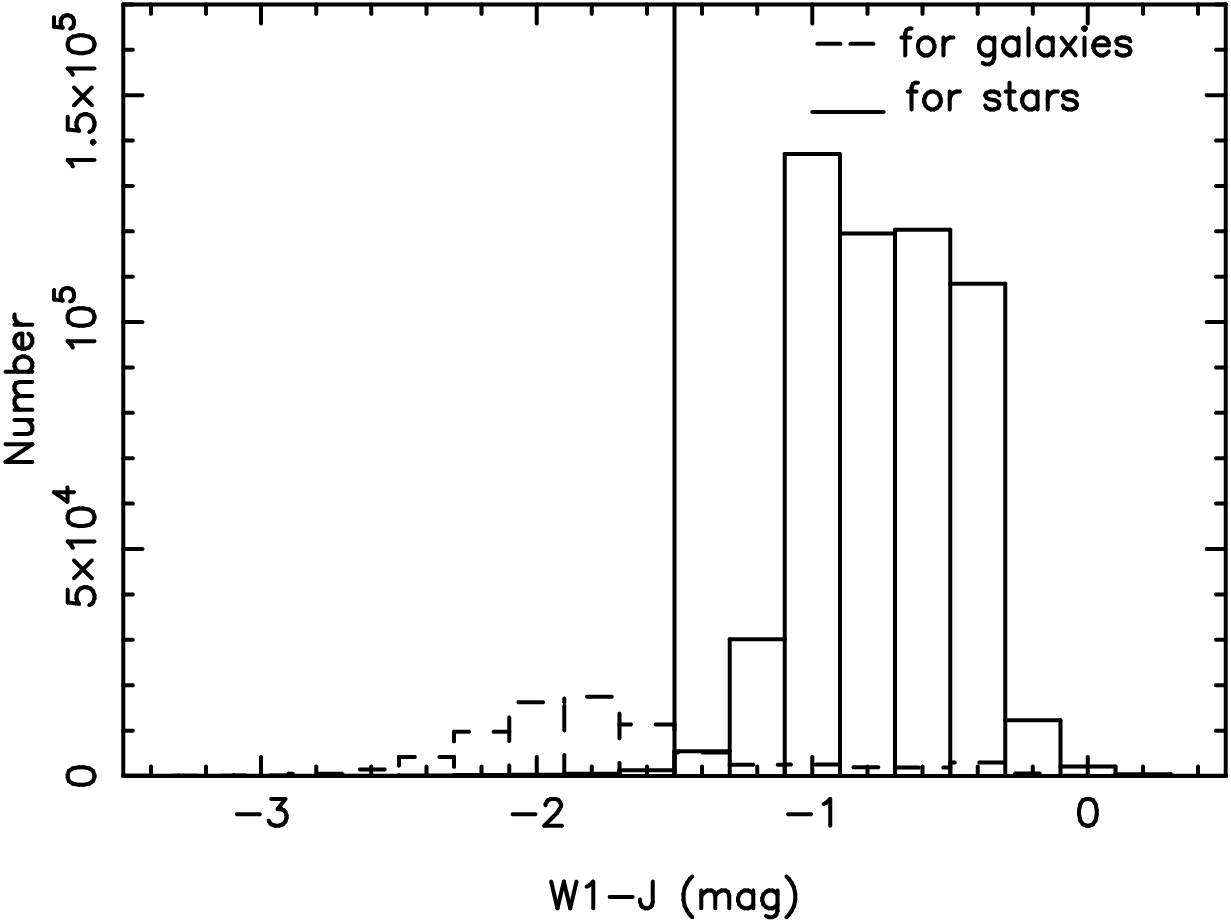}
\end{center}
\vspace*{3mm}
\vspace*{-5mm} {\footnotesize {\bf Figure 1}\quad 
Separation of stars and galaxies by color index, $W1-J$.
The vertical line is $W1-J=-1.5$ mag, which is used as the 
threshold for star-galaxy separation.}
\vspace*{3mm}

To get redshift information for galaxy clusters, we use the catalog of
galaxy photometric
redshifts\footnote{http://surveys.roe.ac.uk/ssa/TWOMPZ} given by
\citet{bjp+14} for 2MASS galaxies. By combining the data of
SuperCOSMOS, 2MASS and WISE, these authors applied an artificial
neural network approach to estimate photometric redshifts of galaxies
covering the whole sky. The photometric redshift has an uncertainty of
$\sigma_z=0.015$ and small percentage of outliers. Note that only
2MASS extended sources have photometric redshifts. The number of
galaxies with photometric redshifts is about one tenth of the galaxies
separated by $W1-J<-1.5$.

Using the WISE-2MASS data together with photometric redshifts by
\citet{bjp+14} in the SDSS Stripe 82 region of $309^{\circ}\leq
\rm{RA} \leq 60^{\circ}$ and $-1.25^{\circ}\leq \rm{Dec} \leq
1.25^{\circ}$, we make cluster identification in the following steps.

For each galaxy with a photometric redshift, we take it as the
temporary central galaxy of a cluster candidate, and the photometric
redshift is taken as the redshift of the cluster candidate. We then
calculate the number of companion galaxies from all separated galaxies
within a projected distance of 1 Mpc. The average number of background
galaxies is estimated using the galaxies within the projected distance
between 2 and 4 Mpc from the assumed central galaxy. We then get the
net number of companion galaxies within 1 Mpc of the central galaxy
after background subtraction, which is taken as a measured
richness, $R_{\rm mea}$. To avoid a cluster identified repeatedly, we
applied the friend-of-friend technique \citep{hg82} to
merge the members into a cluster candidate, and consider only one
cluster candidate within a projected distance of 1 Mpc and a
photometric redshift difference of 0.05. We take the cluster candidate
with the maximum measured richness, $R_{\rm mea}$. 

\vspace*{8mm}
\begin{center}
\includegraphics[angle=0,width=70mm]{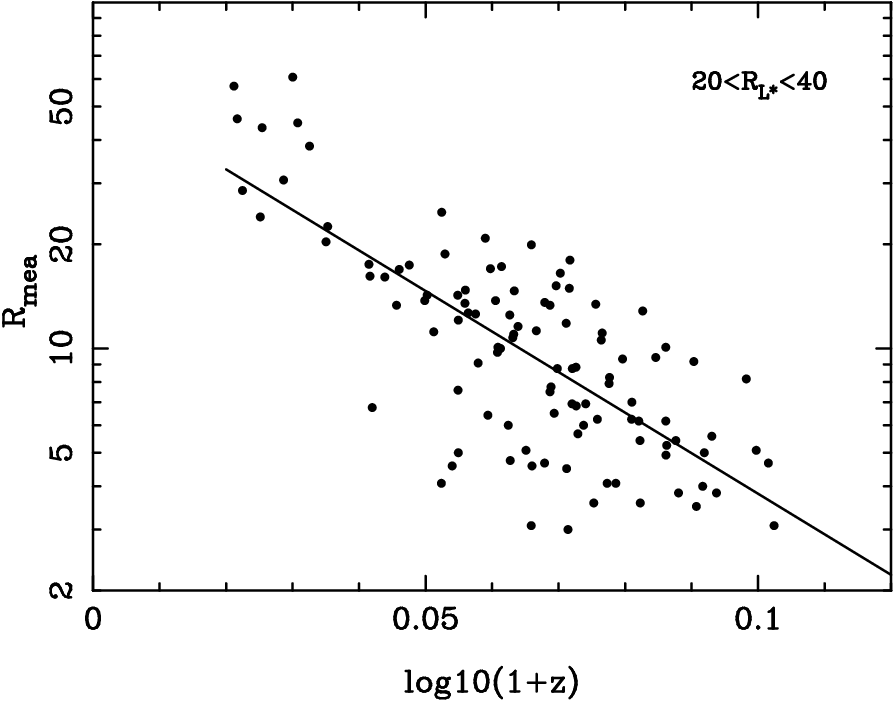}
\end{center}
\vspace*{3mm}
\vspace*{-5mm} {\footnotesize {\bf Figure 2}\quad 
The value of $R_{\rm mea}$ as a function of redshift for the matched
WHL12 clusters of  $20<R_{L*}<40$. The solid line is the best-fit relation.}
\vspace{3mm}

The second step is to find a richness threshold to identify real galaxy
clusters. The best selection is that the cluster richness is related to
cluster mass. Because we use the flux-limited galaxy sample of 
the WISE-2MASS data, the measured richness, $R_{\rm mea}$, strongly 
depends on redshift for clusters with a fixed mass. Here, we define a 
scaled richness with redshift correction,
\begin{equation}
R_{\rm scal}=R_{\rm mea}\,(1+z)^{\alpha},
\end{equation}
to relate cluster mass, where $\alpha$ is the correction slope. Note
that the richness, $R_{L*}$, in the catalog of \citet[][WHL12
  hereafter]{whl12} has a good correlation with cluster mass. Thus, we
cross-match the identified cluster candidates with the WHL12 clusters
of $20<R_{L*}<40$ by using criteria of a projected separation of 1 Mpc
and a redshift difference of 0.05. To get a proper value of $\alpha$,
we plot $R_{\rm scal}$ as a function of redshift for the matched
clusters (Figure~2). The best fit gives $\alpha=9.8\pm1.0$.  We get
similar results when matching WHL12 clusters in other $R_{L*}$
ranges. Figure~3 shows the comparison of $R_{L*}$ with $R_{\rm mea}$
and $R_{\rm scal}$. Clearly, $R_{\rm scal}$ has a better correlation
with $R_{L*}$ than with $R_{\rm mea}$.

We define a cluster to have $R_{\rm scal}\geq30$ in our new
identifications. To avoid the occasional projection effect, we also
require $R_{\rm mea}\geq4$. The cluster candidates with photometric
redshift of $z<0.1$ are excluded because the angular radius varies
rapidly for the fixed radius of 1 Mpc at low redshift, which induces
large uncertainty on richness. Finally, we get 302 clusters in the
redshift range of $0.1<z<0.35$ from the 275 deg$^2$ SDSS Stripe 82
area, which is listed in Table 1.  The histograms for the redshift and
scaled richness are shown in Figure~4. The identified clusters have a
mean redshift of 0.18. If our algorithm is applied to the all sky
SuperCOSMOS, 2MASS and WISE data excluding the Galactic plane of
$|b|>10^{\circ}$, we can find about 37,000 galaxy clusters, which
will greatly enlarge the number of galaxy clusters in the region outside
of the SDSS coverage.

\vspace*{8mm}
\begin{center}
\includegraphics[angle=0,width=70mm]{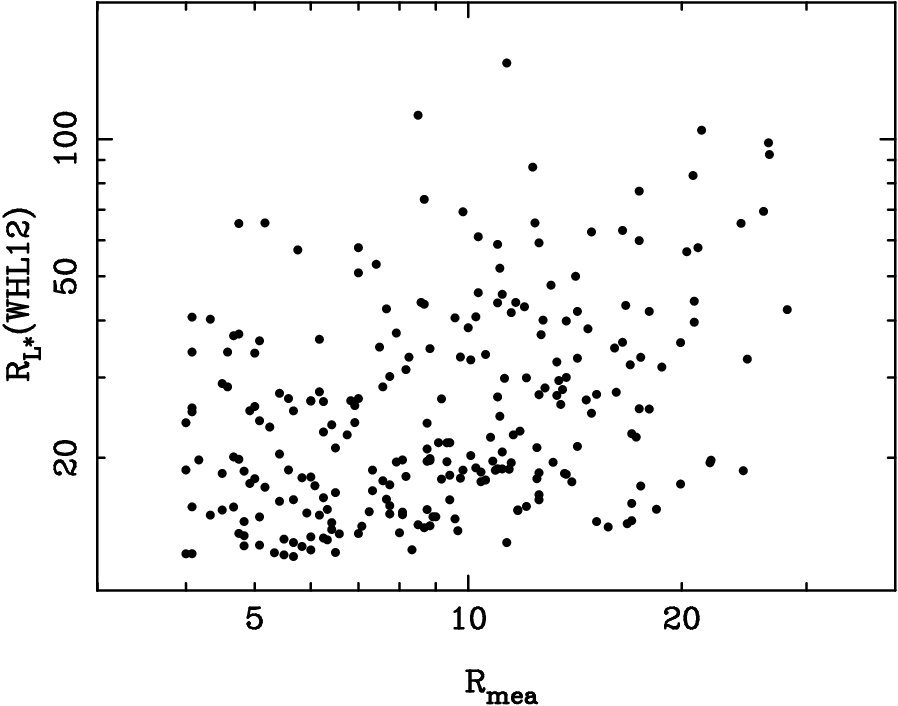}
\includegraphics[angle=0,width=70mm]{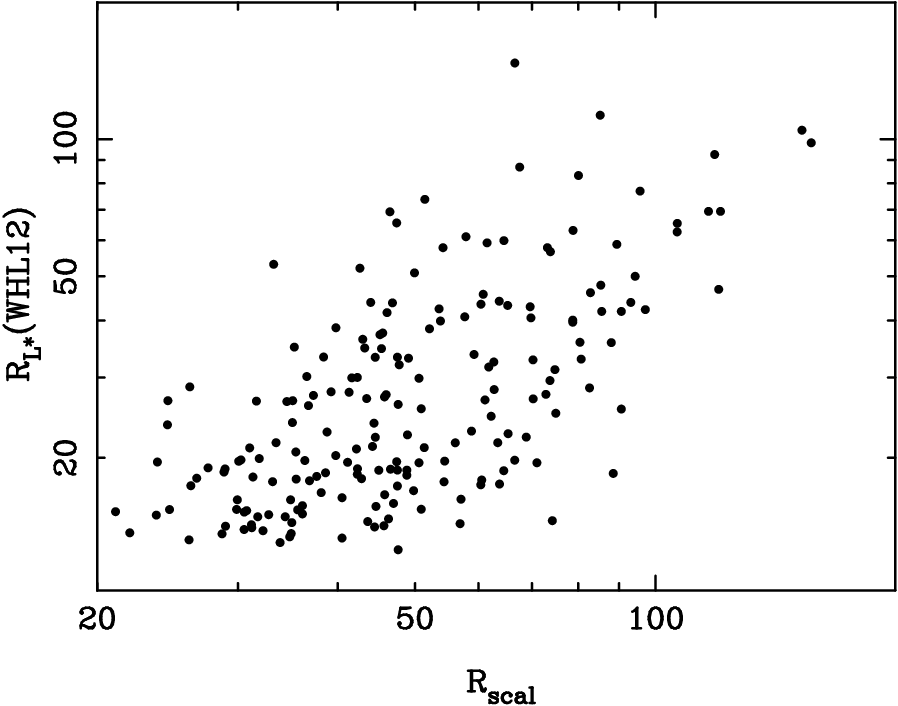}
\end{center}
\vspace*{3mm}
\vspace*{-5mm} {\footnotesize {\bf Figure 3}\quad Comparison of
  measured richness $R_{\rm mea}$ (upper panel) and scaled richness
  $R_{\rm scal}$ (lower panel) with the richness in WHL12 for matched
  cluster candidates.}
\vspace{3mm}

\vspace*{8mm}
\begin{center}
\includegraphics[angle=0,width=70mm]{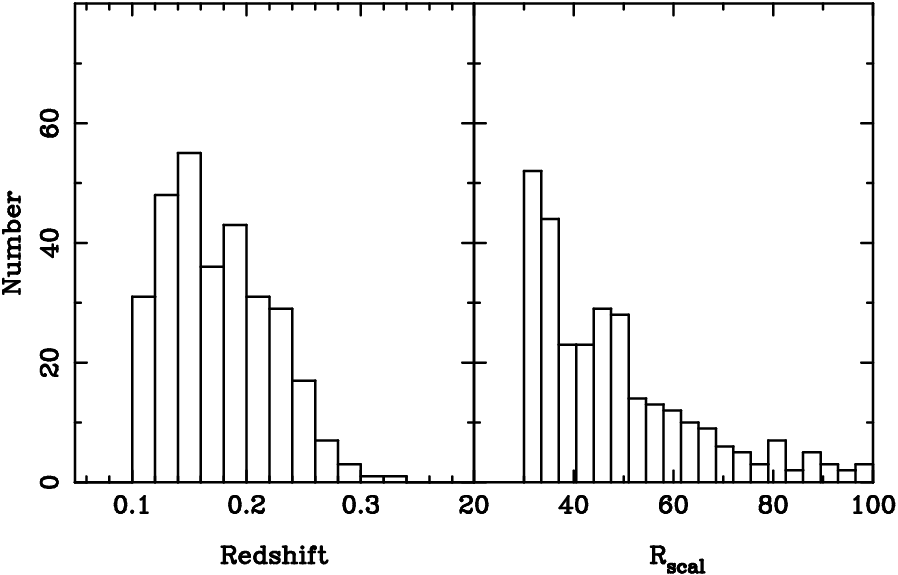}
\end{center}
\vspace*{3mm}
\vspace*{-5mm} {\footnotesize {\bf Figure 4}\quad 
Distributions of redshift and scaled richness for the 302 identified clusters.}
\vspace{3mm}

\vspace*{8mm}
\begin{center}
\includegraphics[angle=0,width=70mm]{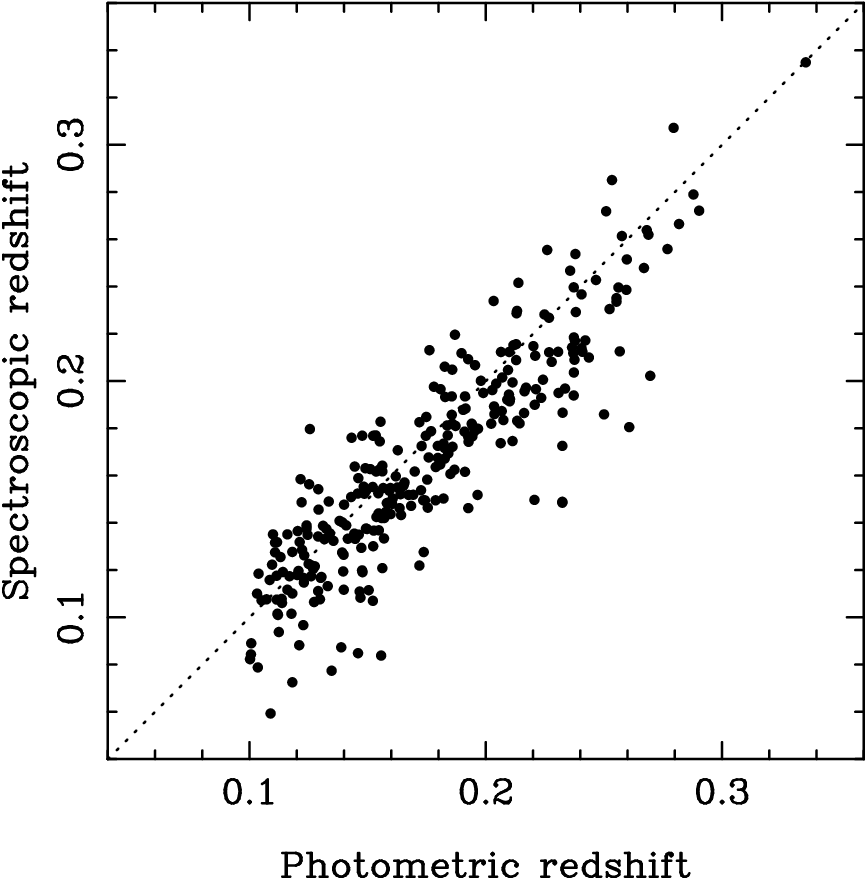}
\end{center}
\vspace*{3mm}
\vspace*{-5mm} {\footnotesize {\bf Figure 5}\quad 
Comparison between cluster photometric redshift and spectroscopic redshift
for 291 clusters.}
\vspace{3mm}

To estimate the uncertainty of photometric redshift of clusters, we
compare the cluster redshifts with the spectroscopic redshifts of the
central galaxies in the SDSS DR7 data \citep{dr7+09}. 291 of
302 central galaxies have their spectroscopic redshifts measured
already. As shown in Figure~5, the cluster photometric redshift is
consistent with spectroscopic redshift with a scatter of 0.022. The
central galaxies without spectroscopic redshifts are not observed by
the SDSS spectroscopic survey, probably due to fiber collision. We
take the value of 0.022 as the typical uncertainty of cluster
photometric redshift.

\section{Comparison with previous galaxy clusters catalogs}

There are many clusters in the Stripe 82 region identified previously
in the catalogs
\citep[i.e.,][]{gsn+02,kma+07,hmk+10,spd+11,gmb11,whl09,whl12}. Generally,
these catalogs have a low false detection rate of $\sim$5\%. The
completeness is as high as $>$90\% for clusters with mass
$>$$1.0\times10^{14}~M_{\odot}$ and is about 50\% for cluster with
mass $0.6\times10^{14}~M_{\odot}$ \citep[e.g.,][]{whl12}.  We regard
the clusters in these catalogs as true clusters and compare them with
302 identified clusters. 247 of 302 (82\%) identified clusters are
matched with the known clusters within a separation of 1 Mpc and a
redshift difference of 0.05. The matched percentage of the identified
clusters with known SDSS clusters varies with richness as expected, as
shown in Figure~6. The matched percentage increases from 82\% with a
richness of $R_{\rm scal} \geq$ 30 to 95\% with a richness of $R_{\rm
  scal} \geq$ 60. Most of the unmatched clusters have a low richness
which all previous methods are less sensitive to detect. In previous
catalogs, the matched percentage between them is in the range of
40\%--80\%, and even lower for poor clusters with a small richness
\citep{spd+11,whl12}. The value of 82\% is very high compared with
previous matched percentage, suggesting that our identification method
is very efficient for finding clusters by using the SuperCOSMOS, 2MASS
and WISE data.

There are 196 clusters matched with the WHL12 clusters, of which 118
(60$\%$) have the central galaxies matched with the brightest
cluster galaxies (BCG) of the WHL12, suggesting that the method
presented in this paper has a high probability to find the BCGs. We
calculate the projected distance between the central galaxies of
identified clusters and BCGs of the WHL12 clusters. As shown in
Figure~7, the distribution is random for the central galaxies 
not matched with the BCGs. For these clusters, the BCGs may not be
located at the positions with the maximum overdensity of galaxy
numbers found within a radius of 1 Mpc. Some of these clusters may
have multiple bright member galaxies in different sub-clusters.

\vspace*{4mm}
\begin{center}
\includegraphics[angle=270,width=70mm]{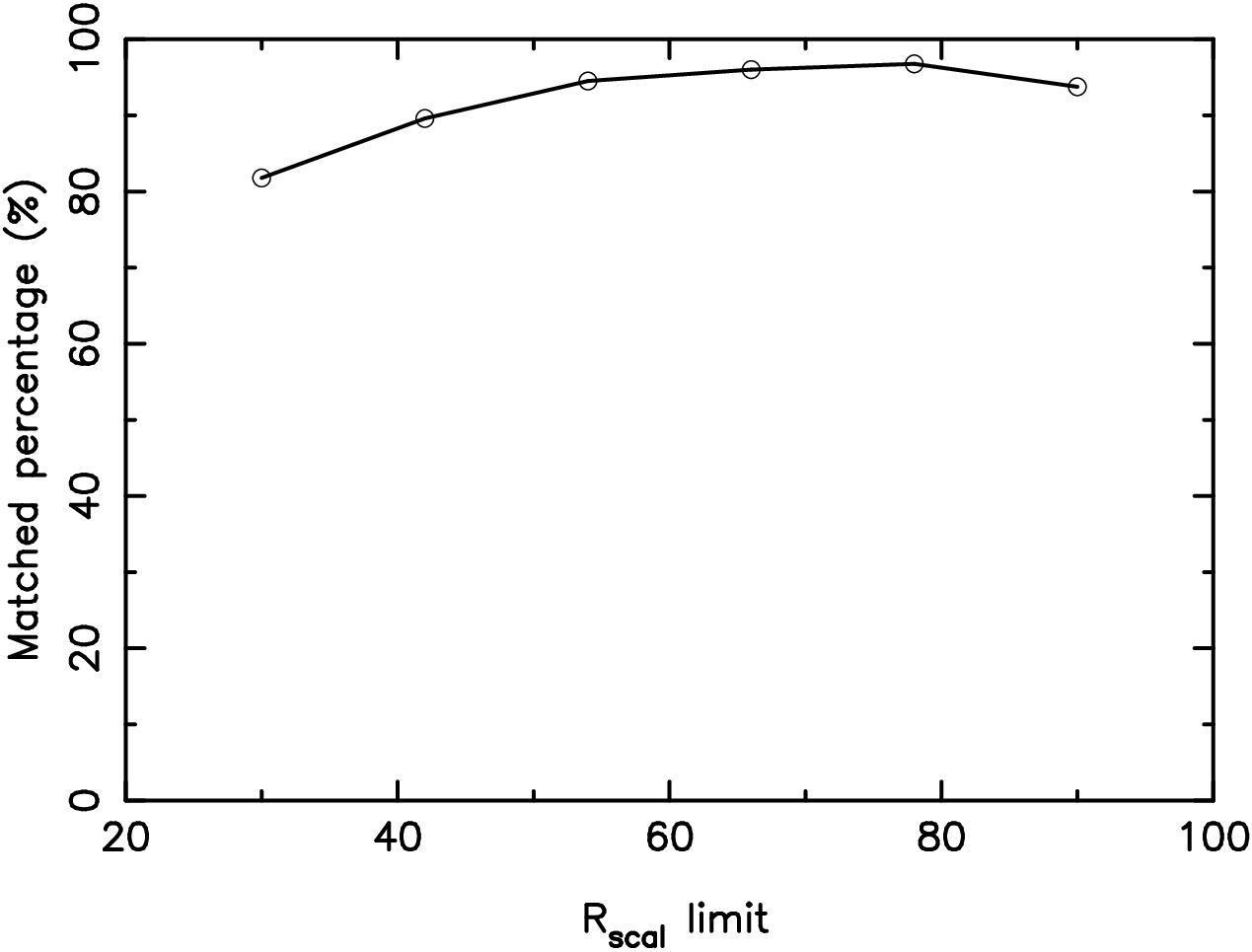}
\end{center}
\vspace*{3mm}
\vspace*{-5mm} {\footnotesize {\bf Figure 6}\quad 
Matched percentage of identified clusters by previously 
known SDSS clusters as a function of richness limit.}
\vspace{3mm}

\vspace*{4mm}
\begin{center}
\includegraphics[angle=0,width=70mm]{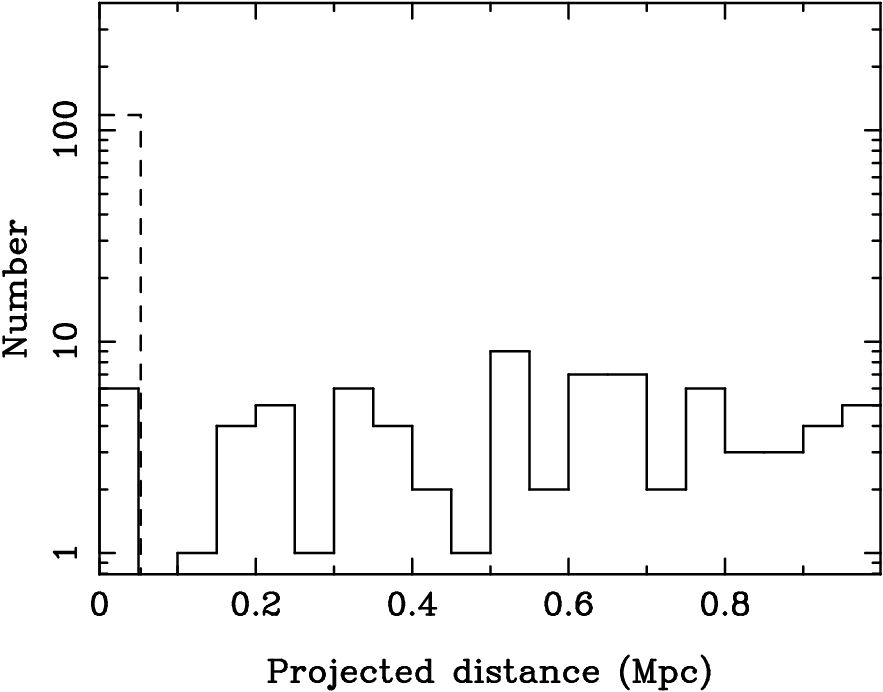}
\end{center}
\vspace*{3mm}
\vspace*{-5mm} {\footnotesize {\bf Figure 7}\quad Distribution of
  projected distance between central galaxies of the identified
  clusters and BCGs of the WHL12 clusters. The dashed line is for the
  matched central galaxies with angular offset less than 3$''$ from
  the BCGs.}
\vspace{3mm}

Many works used red-sequence methods to identify clusters in
multi-band surveys \citep[e.g.,][]{gy00,gy05,kma+07}. The basis of
such methods is that cluster galaxies have similar colors, which are
tightly related to redshift. We check if the colors by the WISE and
2MASS data have tight correlations with redshift for the central
galaxies. We find that the colors, $J-W1$ and $J-W3$, have poor
correlations with redshift (Figure~8). Cluster identification with
$J-W1$ and $J-W3$ may have large uncertainty at redshift of
$z<0.35$. Therefore the photometric redshift data provide a better
opportunity to identify a whole sky galaxy cluster catalog up to
redshift $z\sim0.35$.

\vspace*{4mm}
\begin{center}
\includegraphics[angle=270,width=70mm]{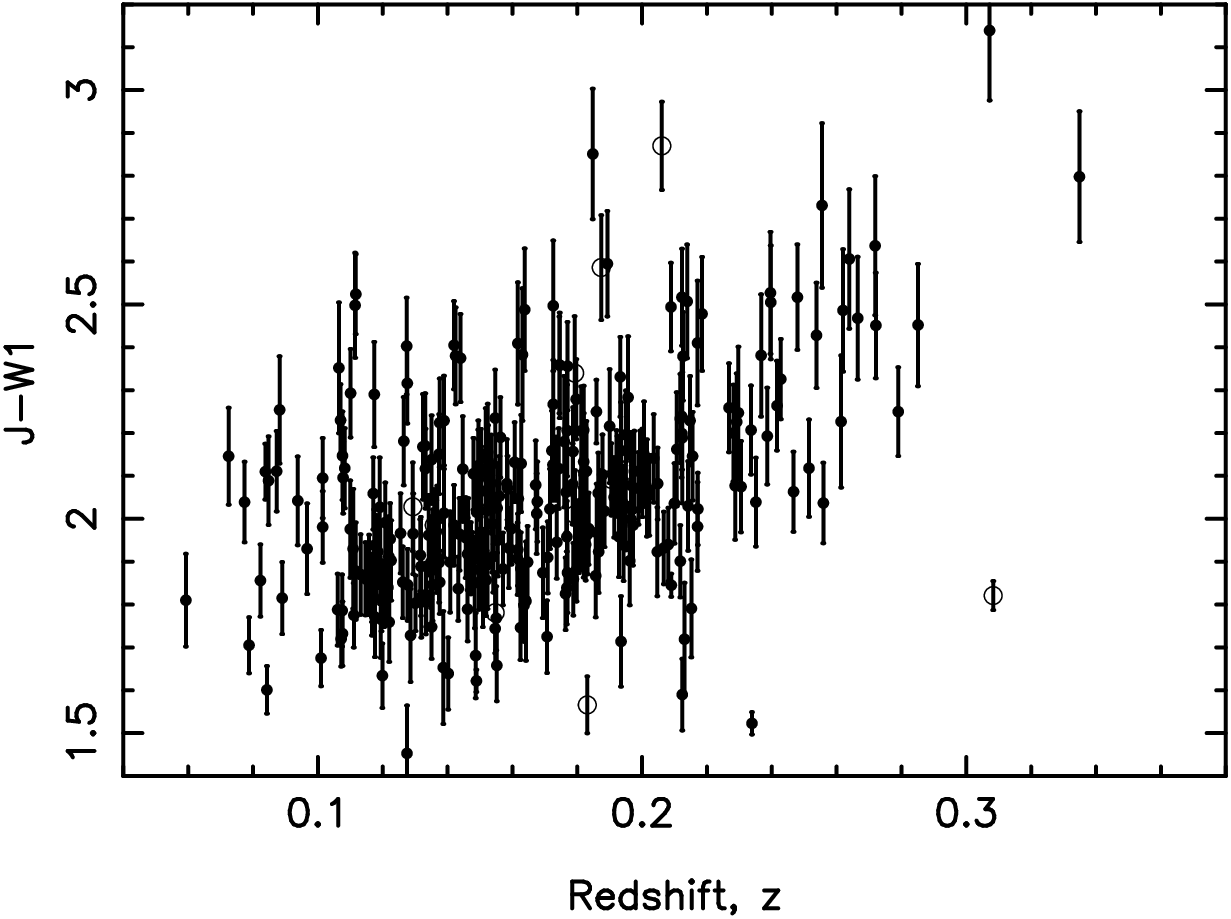}
\includegraphics[angle=270,width=70mm]{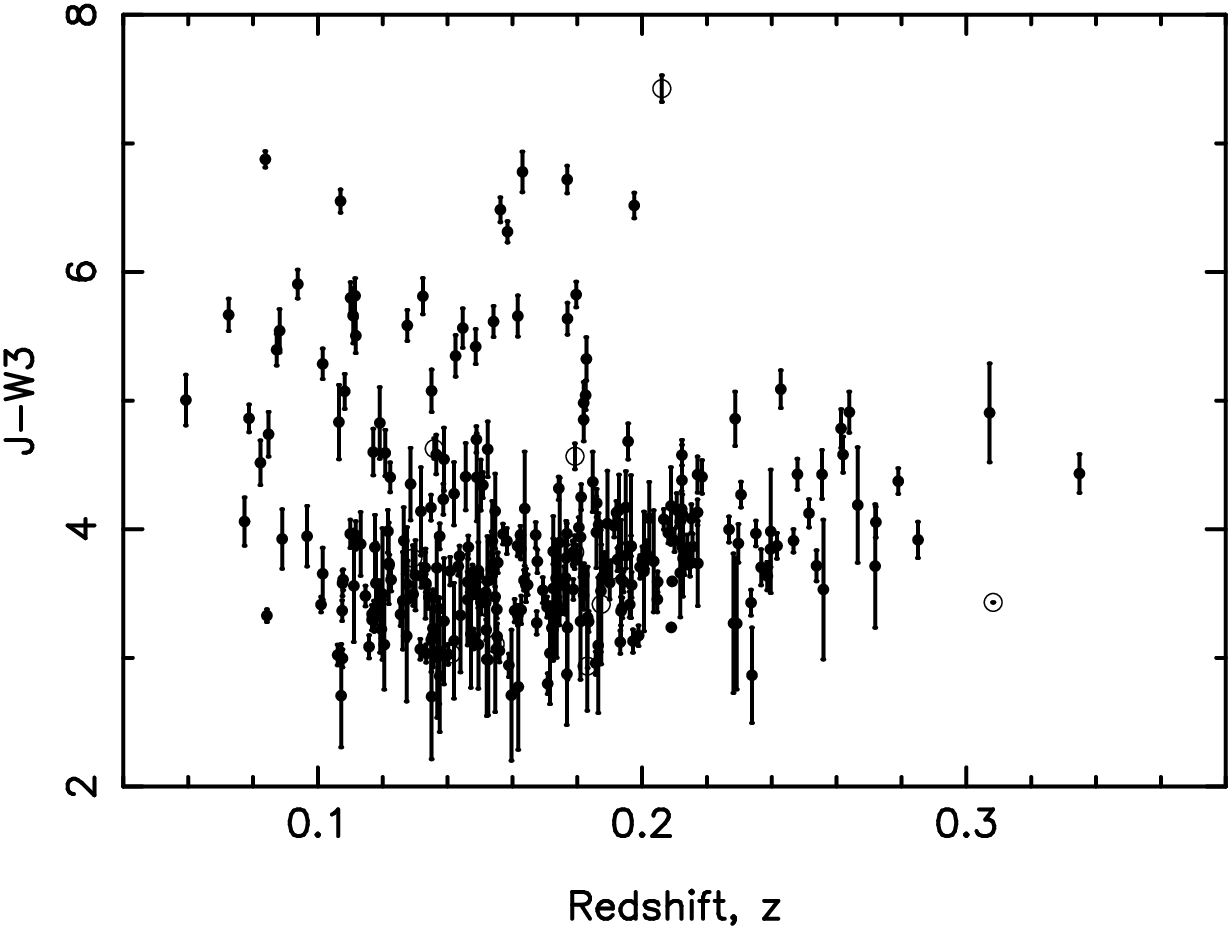}
\end{center}
\vspace*{3mm}
\vspace*{-5mm} {\footnotesize {\bf Figure 8}\quad 
Colors $J-W1$ (upper panel) and $J-W3$ (lower panel) of central galaxies 
versus cluster redshift. Black dots are for spectroscopic
redshifts. Open circles are for photometric redshifts with a typical
uncertainty of 0.022.}
\vspace{3mm}

\section{Summary}

In this paper, we present an efficient method to identify galaxy
clusters by using the SuperCOSMOS, 2MASS and WISE data. First, we
perform star-galaxy separation by color index, $W1-J$. Then, clusters
are identified around the galaxies with estimated photometric
redshift. We get a measured richness and define a scaled richness,
$R_{\rm scal}$, by comparing the richness of \citet{whl12}. Our
method is applied to the data in the SDSS Stripe 82 region and
identify 302 clusters of $R_{\rm scal}\geq30$ in the redshift range of
$0.1<z<0.35$. The photometric redshift has an uncertainty of
0.022. 82\% of our clusters are matched with previous SDSS cluster
catalogs. Our results confirm that this approach has a good potential
to detect many new galaxy clusters in the all sky data of SuperCOSMOS,
2MASS and WISE, especially in the region out of the SDSS coverage.

\acknowledgments

We thank the referee for valuable comments that helped to improve the
paper.  We thank Jun Han for carefully reading the manuscript,
Zhongsheng Yuan and Jun Xu for useful discussion.  The authors are
supported by the National Natural Science Foundation (NNSF) of China
(10833003 and 11103032) and the Young Researcher Grant of National
Astronomical Observatories, Chinese Academy of Sciences.
This publication makes use of data products from the Wide-field Infrared 
Survey Explorer, which is a joint project of the University of California, 
Los Angeles, and the Jet Propulsion Laboratory/California Institute of 
Technology, funded by the National Aeronautics and Space Administration.
This research has made use of the NASA/IPAC Infrared Science Archive, 
which is operated by the Jet Propulsion Laboratory, California Institute 
of Technology, under contract with the National Aeronautics and Space 
Administration.  
Funding for the SDSS and SDSS-II has been provided by the Alfred
P. Sloan Foundation, the Participating Institutions, the National
Science Foundation, the U.S. Department of Energy, the National
Aeronautics and Space Administration, the Japanese Monbukagakusho, the
Max Planck Society, and the Higher Education Funding Council for
England. The SDSS Web site is http://www.sdss.org/.
The SDSS is managed by the Astrophysical Research Consortium for the
Participating Institutions. The Participating Institutions are the
American Museum of Natural History, Astrophysical Institute Potsdam,
University of Basel, Cambridge University, Case Western Reserve
University, University of Chicago, Drexel University, Fermilab, the
Institute for Advanced Study, the Japan Participation Group, Johns
Hopkins University, the Joint Institute for Nuclear Astrophysics, the
Kavli Institute for Particle Astrophysics and Cosmology, the Korean
Scientist Group, the Chinese Academy of Sciences (LAMOST), Los Alamos
National Laboratory, the Max Planck Institute for Astronomy (MPIA),
the Max Planck Institute for Astrophysics (MPA), New Mexico State
University, Ohio State University, University of Pittsburgh,
University of Portsmouth, Princeton University, the United States
Naval Observatory, and the University of Washington.

\begin{table}[htbp]
  \centering
\footnotesize
    \caption{The 302 identified Cluster Candidates in the Stripe 82}
    \begin{tabular}{ c c c c c c c c c c}
      \hline
      \hline
   ID &    R.A.      &   Decl.     & $z_p$     &  $z_s$      &  $J$    &  $W1$   & $R_{scal}$ & $R_{mea}$  &   Note   \\   
      &   (deg)      &   (deg)     &           &             &  (mag)  & (mag)   &            &            & \\  
  (1) &    (2)       &   (3)       &   (4)     &  (5)        &    (6)  &    (7)	 &    (8)     & (9)        &  (10)      \\      
      \hline 
     1   &   312.7079 &    -1.0438 &     0.1335 &     0.1490 &  15.53 $\pm$ 0.10 &  13.52 $\pm$ 0.03 &  30.03 &   9.08 &  Geach \\
     2   &   312.8900 &    -0.0495 &     0.1400 &     0.1477 &  15.15 $\pm$ 0.06 &  13.25 $\pm$ 0.02 &  41.98 &  12.08 &  MaxBCG,AMF,Geach,WHL12 \\
     3   &   313.0400 &     0.3102 &     0.1754 &     0.1464 &  15.55 $\pm$ 0.09 &  13.60 $\pm$ 0.03 &  33.57 &   7.42 &  MaxBCG,GMBCG,Geach,WHL12 \\
     4   &   314.3147 &     1.2120 &     0.1788 &     0.1636 &  15.61 $\pm$ 0.08 &  13.81 $\pm$ 0.03 &  36.30 &   7.83  & New \\ 
     5   &   315.0705 &     0.4328 &     0.1035 &     0.0788 &  15.39 $\pm$ 0.06 &  13.69 $\pm$ 0.03 &  40.71 &  15.50  & New \\ 
     6   &   315.7259 &     0.8867 &     0.2576 &     0.2614 &  16.45 $\pm$ 0.15 &  14.22 $\pm$ 0.04 &  83.93 &  10.33 &  AMF,Geach,WHL12 \\
     7   &   315.7609 &     0.8942 &     0.1340 &     0.1355 &  15.07 $\pm$ 0.07 &  13.12 $\pm$ 0.03 &  62.27 &  18.75 &  GMBCG,Geach,WHL12 \\
     8   &   315.8023 &    -1.0171 &     0.1912 &     0.1616 &  15.64 $\pm$ 0.08 &  13.73 $\pm$ 0.03 &  75.73 &  14.92 &  GMBCG,AMF,Geach,WHL12 \\
     9   &   315.8495 &     0.9987 &     0.1477 &     0.1769 &  16.06 $\pm$ 0.10 &  13.85 $\pm$ 0.04 &  36.20 &   9.83 &  WHL12 \\
    10   &   316.1830 &     0.9142 &     0.1826 &     0.1672 &  15.88 $\pm$ 0.10 &  13.80 $\pm$ 0.03 &  36.94 &   7.75 &  GMBCG,AMF,Geach,WHL12 \\
    11   &   317.9637 &     0.0149 &     0.2209 &     0.2107 &  16.18 $\pm$ 0.13 &  14.02 $\pm$ 0.03 &  43.97 &   7.00 &  MaxBCG,GMBCG,AMF,Geach,WHL12 \\
    12   &   318.5470 &    -0.5487 &     0.2553 &     0.2351 &  15.95 $\pm$ 0.10 &  13.91 $\pm$ 0.03 &  31.99 &   4.00 &  AMF,WHL12 \\
    13   &   319.3704 &     0.1176 &     0.2553 &     0.2336 &  16.05 $\pm$ 0.10 &  13.84 $\pm$ 0.03 &  33.35 &   4.17 &  MaxBCG,GMBCG,AMF,Geach,WHL12 \\
    14   &   319.5089 &     0.8675 &     0.1562 &     0.1208 &  15.72 $\pm$ 0.09 &  13.73 $\pm$ 0.03 &  36.30 &   9.25 &  Geach \\
    15   &   319.9571 &     1.1862 &     0.1204 &     0.1365 &  15.52 $\pm$ 0.07 &  13.55 $\pm$ 0.02 &  49.88 &  16.67 &  AMF \\
    16   &   320.9352 &    -0.6985 &     0.2308 &     0.1950 &  15.89 $\pm$ 0.10 &  13.69 $\pm$ 0.03 &  49.39 &   7.33 &  MaxBCG,GMBCG,AMF,Geach,WHL12 \\
    17   &   321.1893 &    -0.5187 &     0.2524 &     0.2305 &  15.98 $\pm$ 0.10 &  13.90 $\pm$ 0.04 &  54.87 &   7.00 &  MaxBCG,GMBCG,AMF,Geach,WHL12 \\
    18   &   321.4364 &     1.0129 &     0.1497 &     0.1374 &  15.54 $\pm$ 0.09 &  13.64 $\pm$ 0.03 &  65.10 &  17.42 &  GMBCG,AMF,Geach,WHL12 \\
    19   &   322.4165 &     0.0891 &     0.2033 &     0.2339 &  14.98 $\pm$ 9.99 &  13.46 $\pm$ 0.03 &  68.29 &  12.33 &  MaxBCG,GMBCG,Geach,WHL12 \\
    20   &   322.4345 &     0.3405 &     0.1952 &     0.1787 &  15.68 $\pm$ 0.10 &  13.52 $\pm$ 0.03 &  46.15 &   8.83 &  MaxBCG,Geach,WHL12 \\
    21   &   322.4928 &    -0.3298 &     0.1547 &       ...  &  15.68 $\pm$ 0.09 &  13.90 $\pm$ 0.03 &  42.69 &  11.00 &  MaxBCG,GMBCG,Geach,WHL12 \\
    22   &   322.5165 &    -0.3523 &     0.2405 &     0.2367 &  16.37 $\pm$ 0.14 &  13.99 $\pm$ 0.03 &  50.52 &   7.00 &  MaxBCG,GMBCG,Geach,WHL12 \\
    23   &   322.5982 &    -0.0492 &     0.1569 &     0.1334 &  15.64 $\pm$ 0.07 &  13.82 $\pm$ 0.03 &  54.24 &  13.75 &  MaxBCG,GMBCG,Geach,WHL12 \\
    24   &   323.1762 &     1.1921 &     0.2061 &       ...  &  16.11 $\pm$ 0.10 &  13.24 $\pm$ 0.02 &  37.72 &   6.67  & New \\ 
    25   &   323.5251 &    -0.5225 &     0.2381 &     0.2292 &  15.78 $\pm$ 0.11 &  13.55 $\pm$ 0.03 &  53.79 &   7.58 &  GMBCG,Geach \\
    26   &   323.5826 &     0.5438 &     0.1967 &     0.1798 &  15.61 $\pm$ 0.07 &  13.73 $\pm$ 0.03 &  41.37 &   7.83 &  GMBCG,Geach \\
    27   &   323.8004 &    -1.0496 &     0.3083 &       ...  &  15.67 $\pm$ 9.99 &  13.85 $\pm$ 0.03 &  67.92 &   5.92 &  GMBCG,Geach,WHL12 \\
    28   &   323.9428 &     0.1158 &     0.1169 &     0.1174 &  15.81 $\pm$ 0.12 &  13.52 $\pm$ 0.03 & 120.70 &  41.42 &  Abell,MaxBCG,GMBCG,AMF,Geach,WHL12 \\
    29   &   324.6582 &    -0.3506 &     0.1521 &     0.1552 &  15.24 $\pm$ 0.07 &  13.58 $\pm$ 0.05 &  30.44 &   8.00  & New \\ 
    30   &   326.1441 &     1.1394 &     0.2267 &     0.2268 &  16.00 $\pm$ 0.10 &  13.74 $\pm$ 0.03 &  89.52 &  13.67 &  Abell,Geach,WHL12 \\
    31   &   326.3670 &    -0.7800 &     0.1850 &     0.1607 &  15.36 $\pm$ 0.09 &  13.23 $\pm$ 0.02 &  73.60 &  15.17 &  MaxBCG,GMBCG,Geach,WHL12 \\
    32   &   326.5361 &    -0.2424 &     0.2130 &     0.2287 &  16.15 $\pm$ 0.12 &  14.07 $\pm$ 0.04 &  31.67 &   5.33 &  Geach,WHL12 \\
    33   &   326.8617 &     0.7289 &     0.1181 &     0.0725 &  15.72 $\pm$ 0.11 &  13.57 $\pm$ 0.03 &  51.21 &  17.42 &  Abell,MaxBCG,Geach,WHL12 \\
    34   &   328.5980 &     0.0844 &     0.1603 &     0.1480 &  15.05 $\pm$ 0.08 &  12.94 $\pm$ 0.03 &  46.50 &  11.50 &  MaxBCG,GMBCG,Geach,WHL12 \\
    35   &   328.6150 &     0.6436 &     0.1831 &       ...  &  14.86 $\pm$ 0.06 &  13.29 $\pm$ 0.03 &  31.11 &   6.50 &  Abell,MaxBCG,GMBCG,AMF,Geach,WHL12 \\
    36   &   328.7676 &     0.8724 &     0.2064 &     0.2123 &  16.14 $\pm$ 0.11 &  13.62 $\pm$ 0.03 &  58.51 &  10.33 &  GMBCG,AMF,Geach,WHL12 \\
    37   &   328.8372 &     1.1254 &     0.2370 &     0.2118 &  15.67 $\pm$ 0.08 &  13.77 $\pm$ 0.03 &  39.91 &   5.67 &  Geach,WHL12 \\
    38   &   328.9167 &     0.5372 &     0.1858 &     0.2048 &  15.51 $\pm$ 0.08 &  13.43 $\pm$ 0.03 &  46.75 &   9.58 &  MaxBCG,AMF,Geach,WHL12 \\
    39   &   329.3690 &    -0.9485 &     0.1273 &     0.1065 &  16.39 $\pm$ 0.15 &  14.04 $\pm$ 0.03 &  32.08 &  10.17  & New \\ 
    40   &   329.3874 &    -0.9288 &     0.2372 &     0.1939 &  15.74 $\pm$ 0.08 &  13.70 $\pm$ 0.03 &  71.06 &  10.08 &  MaxBCG,Geach,WHL12 \\
    41   &   329.9931 &    -0.6349 &     0.1108 &     0.1275 &  15.54 $\pm$ 0.10 &  14.09 $\pm$ 0.05 &  47.27 &  17.00 &  AMF,WHL12 \\
    42   &   330.1559 &    -0.5459 &     0.1398 &     0.1265 &  15.44 $\pm$ 0.10 &  13.26 $\pm$ 0.02 &  49.43 &  14.25 &  MaxBCG,Geach,WHL12 \\
    43   &   331.2684 &    -0.5626 &     0.1597 &     0.1437 &  15.73 $\pm$ 0.08 &  13.77 $\pm$ 0.03 &  69.46 &  17.25 &  MaxBCG,Geach,WHL12 \\
    44   &   331.6891 &     1.0279 &     0.2372 &     0.2397 &  16.25 $\pm$ 0.13 &  13.74 $\pm$ 0.03 &  43.50 &   6.17 &  AMF,Geach,WHL12 \\
    45   &   332.3338 &     1.2364 &     0.1606 &     0.1503 &  15.64 $\pm$ 0.09 &  13.67 $\pm$ 0.03 &  37.50 &   9.25  & New \\ 
    46   &   334.1963 &    -0.9939 &     0.1965 &     0.1518 &  15.78 $\pm$ 0.11 &  13.64 $\pm$ 0.03 &  31.64 &   6.00 &  WHL12 \\
    47   &   335.4678 &    -0.9728 &     0.3354 &     0.3349 &  16.51 $\pm$ 0.15 &  13.71 $\pm$ 0.03 &  65.23 &   4.75 &  Geach \\
    48   &   335.4820 &    -1.0553 &     0.1051 &     0.1072 &  14.89 $\pm$ 0.06 &  13.17 $\pm$ 0.02 &  64.94 &  24.42 &  WHL12 \\
    49   &   335.5464 &    -1.0006 &     0.2206 &     0.1497 &  15.80 $\pm$ 0.09 &  13.92 $\pm$ 0.03 &  44.97 &   7.17  & New \\ 
    50   &   335.6234 &     0.5694 &     0.1859 &     0.1722 &  15.53 $\pm$ 0.09 &  13.37 $\pm$ 0.02 &  67.16 &  13.75 &  Abell,MaxBCG,Geach,WHL12 \\
    51   &   335.7316 &     0.5291 &     0.1445 &     0.1638 &  16.41 $\pm$ 0.14 &  13.92 $\pm$ 0.03 &  45.54 &  12.67 &  Abell,GMBCG,AMF,Geach,WHL12 \\
    52   &   336.0885 &     0.3597 &     0.1570 &     0.1420 &  15.71 $\pm$ 0.09 &  13.73 $\pm$ 0.03 &  65.81 &  16.67 &  MaxBCG,GMBCG,Geach,WHL12 \\
    53   &   336.2299 &    -0.3840 &     0.1559 &     0.1420 &  15.38 $\pm$ 0.10 &  12.98 $\pm$ 0.02 &  54.80 &  14.00 &  MaxBCG,Geach,WHL12 \\
    54   &   337.2881 &     0.4193 &     0.1523 &     0.1301 &  15.41 $\pm$ 0.06 &  13.61 $\pm$ 0.03 &  30.80 &   8.08 &  Geach,WHL12 \\
    55   &   337.5326 &    -0.0037 &     0.2279 &     0.2081 &  15.72 $\pm$ 0.08 &  13.78 $\pm$ 0.03 &  86.35 &  13.08 &  MaxBCG,GMBCG,AMF,Geach,WHL12 \\
    56   &   337.8498 &     0.2527 &     0.2769 &     0.2559 &  15.97 $\pm$ 0.09 &  13.93 $\pm$ 0.03 &  75.82 &   8.17 &  MaxBCG,GMBCG,Geach,WHL12 \\
    57   &   337.9117 &    -1.0345 &     0.1458 &     0.1524 &  15.94 $\pm$ 0.11 &  13.78 $\pm$ 0.03 &  31.46 &   8.67  & New \\ 
    58   &   338.4055 &     0.6970 &     0.1742 &     0.1495 &  15.99 $\pm$ 0.11 &  13.87 $\pm$ 0.03 &  56.38 &  12.58 &  AMF \\
    59   &   338.5464 &     1.0411 &     0.1905 &       ...  &  15.85 $\pm$ 0.08 &  13.76 $\pm$ 0.03 &  60.60 &  12.00 &  MaxBCG \\
    60   &   338.8853 &    -1.1846 &     0.1007 &     0.0890 &  15.34 $\pm$ 0.08 &  13.52 $\pm$ 0.03 &  41.57 &  16.17 &  GMBCG,Geach,WHL12 \\
    61   &   339.5265 &     0.5325 &     0.2373 &     0.2036 &  15.89 $\pm$ 0.10 &  13.75 $\pm$ 0.03 &  34.71 &   4.92 &  WHL12 \\
    62   &   339.5725 &    -1.0250 &     0.1301 &     0.1167 &  15.43 $\pm$ 0.07 &  13.60 $\pm$ 0.03 &  64.22 &  19.92 &  MaxBCG,WHL12 \\
    63   &   339.6037 &     0.4290 &     0.1476 &     0.1199 &  15.58 $\pm$ 0.07 &  13.95 $\pm$ 0.03 &  36.79 &  10.00 &  GMBCG \\                    
    64   &   339.6763 &    -0.4495 &     0.1737 &     0.1276 &  15.82 $\pm$ 0.09 &  13.50 $\pm$ 0.03 &  44.67 &  10.00  & New \\ 
    65   &   339.7389 &     0.9962 &     0.1241 &     0.1389 &  15.87 $\pm$ 0.10 &  13.86 $\pm$ 0.03 &  41.59 &  13.50  & New \\ 
        \hline                                                                           
        \hline                                                                          
    \end{tabular}                                                                       
    \small                                                                              
\end{table}

\addtocounter{table}{-1}                                                                
                                                                                        
\begin{table}[htbp]                                                                    
  \centering                                                                             
\footnotesize                                                                          
    \caption{continued}                                                                  
    \begin{tabular}{ c c c c c c c c c c}                                               
      \hline                                                                            
      \hline                                                                            
   ID &    R.A.      &   Decl.     & $z_p$     &  $z_s$      &  $J$    &  $W1$   & $R_{scal}$ & $R_{mea}$  &   Note   \\   
      &   (deg)      &   (deg)     &           &             &  (mag)  & (mag)   &          &            & \\  
  (1) &    (2)       &   (3)       &   (4)     &  (5)        &    (6)  &    (7)	 &    (8)   & (9)        &  (10)      \\      
      \hline                    
    66   &   339.7720 &     0.6534 &     0.2375 &     0.2090 &  15.60            &  13.76 $\pm$ 0.03 &  54.20 &   7.67 &  MaxBCG,Geach,WHL12 \\
    67   &   339.9143 &     1.1757 &     0.1641 &     0.1433 &  15.70 $\pm$ 0.07 &  13.86 $\pm$ 0.03 &  45.42 &  10.92 &  Geach,WHL12 \\
    68   &   340.6530 &     0.8837 &     0.1222 &     0.1286 &  15.65 $\pm$ 0.10 &  13.92 $\pm$ 0.04 &  50.59 &  16.67  & New \\ 
    69   &   340.8411 &     0.3386 &     0.1089 &     0.0593 &  15.78 $\pm$ 0.10 &  13.97 $\pm$ 0.04 &  46.78 &  17.08 &  WHL12 \\
    70   &   341.1638 &     0.5700 &     0.1137 &     0.1061 &  15.02 $\pm$ 0.08 &  13.23 $\pm$ 0.02 &  44.77 &  15.75  & New \\ 
    71   &   341.7559 &     0.9144 &     0.1793 &       ...  &  15.81 $\pm$ 0.10 &  13.80 $\pm$ 0.03 &  34.92 &   7.50 &  MaxBCG,Geach \\
    72   &   342.0809 &    -0.6117 &     0.2099 &     0.2123 &  15.90 $\pm$ 0.08 &  13.70 $\pm$ 0.03 &  46.02 &   7.92 &  GMBCG,AMF,WHL12 \\
    73   &   342.1753 &     0.9131 &     0.1359 &       ...  &  15.13 $\pm$ 0.08 &  13.15 $\pm$ 0.03 &  38.73 &  11.50 &  MaxBCG,AMF \\
    74   &   342.5280 &     0.8538 &     0.1927 &     0.1743 &  15.68 $\pm$ 0.09 &  13.56 $\pm$ 0.03 &  47.87 &   9.33 &  MaxBCG,GMBCG,AMF,Geach,WHL12 \\
    75   &   342.7946 &    -0.7928 &     0.2365 &     0.2142 &  15.92 $\pm$ 0.10 &  13.89 $\pm$ 0.03 &  38.59 &   5.50 &  Geach,WHL12 \\
    76   &   342.9160 &     1.1677 &     0.1388 &     0.0873 &  15.65 $\pm$ 0.09 &  13.54 $\pm$ 0.03 &  30.98 &   9.00  & New \\ 
    77   &   343.9557 &     0.5725 &     0.1910 &     0.1786 &  15.85 $\pm$ 0.09 &  13.77 $\pm$ 0.03 &  35.07 &   6.92  & New \\ 
    78   &   344.0618 &    -0.5811 &     0.1032 &     0.1100 &  15.93 $\pm$ 0.11 &  13.95 $\pm$ 0.03 &  81.27 &  31.00 &  Abell,MaxBCG,GMBCG,AMF,Geach,WHL12 \\
    79   &   344.1515 &    -0.4658 &     0.1469 &     0.1083 &  15.55 $\pm$ 0.09 &  13.43 $\pm$ 0.03 &  74.41 &  20.33 &  Abell,MaxBCG,GMBCG,AMF,Geach,WHL12 \\
    80   &   344.3402 &    -1.1362 &     0.1792 &       ...  &  15.82 $\pm$ 0.13 &  13.48 $\pm$ 0.03 &  38.00 &   8.17 &  MaxBCG,Geach,WHL12 \\
    81   &   344.5970 &     0.2678 &     0.1291 &     0.1542 &  15.89 $\pm$ 0.09 &  13.85 $\pm$ 0.03 &  57.84 &  18.08 &  MaxBCG,WHL12 \\
    82   &   344.6125 &    -0.1018 &     0.1872 &     0.1810 &  15.72 $\pm$ 0.09 &  13.74 $\pm$ 0.03 &  54.62 &  11.08 &  MaxBCG,Geach \\
    83   &   344.6595 &     0.5849 &     0.1252 &     0.1563 &  15.67 $\pm$ 0.09 &  13.48 $\pm$ 0.03 &  32.60 &  10.50  & New \\ 
    84   &   344.7387 &     1.2057 &     0.1085 &     0.1158 &  15.38 $\pm$ 0.09 &  13.51 $\pm$ 0.03 &  47.80 &  17.50  & New \\ 
    85   &   345.0132 &     1.1510 &     0.1304 &     0.1171 &  15.73 $\pm$ 0.08 &  13.67 $\pm$ 0.03 &  40.63 &  12.58  & New \\ 
    86   &   345.0288 &     0.2221 &     0.1874 &       ...  &  15.62 $\pm$ 0.12 &  13.03 $\pm$ 0.02 &  62.11 &  12.58 &  GMBCG,AMF,Geach,WHL12 \\
    87   &   345.0460 &     0.3640 &     0.2248 &     0.2282 &  15.70 $\pm$ 0.11 &  13.50 $\pm$ 0.03 &  35.00 &   5.42 &  WHL12 \\
    88   &   345.5182 &     0.2212 &     0.1293 &       ...  &  15.79 $\pm$ 0.10 &  13.76 $\pm$ 0.03 &  33.39 &  10.42 &  MaxBCG,GMBCG,Geach,WHL12 \\
    89   &   345.9451 &     0.7804 &     0.1566 &     0.1547 &  15.84 $\pm$ 0.11 &  13.60 $\pm$ 0.03 &  39.66 &  10.08 &  MaxBCG \\
    90   &   348.1483 &     0.1520 &     0.1260 &     0.1174 &  15.03 $\pm$ 0.08 &  13.09 $\pm$ 0.03 &  44.52 &  14.25 &  MaxBCG,Geach,WHL12 \\
    91   &   350.2303 &     0.5441 &     0.2067 &     0.1873 &  15.73 $\pm$ 0.09 &  13.71 $\pm$ 0.03 &  62.91 &  11.08 &  MaxBCG,Geach,WHL12 \\
    92   &   350.6064 &     1.0693 &     0.1396 &     0.1194 &  15.09 $\pm$ 0.08 &  13.32 $\pm$ 0.04 &  30.30 &   8.75 &  CE,MaxBCG,AMF,Geach,WHL12 \\
    93   &   350.7556 &     0.8997 &     0.1269 &     0.1205 &  15.21 $\pm$ 0.07 &  13.39 $\pm$ 0.03 &  41.43 &  13.17 &  MaxBCG,AMF,WHL12 \\
    94   &   351.0899 &     0.3193 &     0.1735 &     0.1495 &  15.11 $\pm$ 0.08 &  12.99 $\pm$ 0.03 &  88.83 &  19.92 &  Abell,CE,MaxBCG,GMBCG,AMF,Geach,WHL12 \\
    95   &   351.1997 &     0.9442 &     0.1038 &     0.1185 &  15.22 $\pm$ 0.08 &  13.30 $\pm$ 0.03 &  30.94 &  11.75 &  CE,WHL12 \\
    96   &   354.4262 &     0.3038 &     0.1207 &     0.1196 &  15.17 $\pm$ 0.07 &  13.36 $\pm$ 0.03 &  77.26 &  25.75 &  Abell \\
    97   &   355.0894 &     0.2696 &     0.1447 &     0.1332 &  15.38 $\pm$ 0.09 &  13.26 $\pm$ 0.02 &  32.41 &   9.00 &  CE \\
    98   &   355.2489 &     0.0817 &     0.1748 &     0.1848 &  15.66 $\pm$ 0.15 &  12.81 $\pm$ 0.03 & 119.64 &  26.58 &  Abell,CE,MaxBCG,GMBCG,AMF,Geach,WHL12 \\
    99   &   355.9156 &     0.4243 &     0.2500 &     0.1859 &  15.66 $\pm$ 0.07 &  13.41 $\pm$ 0.02 &  50.10 &   6.50  & New \\ 
   100   &   356.0726 &    -0.8548 &     0.2607 &     0.1805 &  15.87 $\pm$ 0.08 &  13.91 $\pm$ 0.03 &  33.87 &   4.08  & New \\ 
   101   &   356.5196 &    -0.1857 &     0.2818 &     0.2665 &  16.45 $\pm$ 0.14 &  13.98 $\pm$ 0.03 &  48.76 &   5.08 &  CE,MaxBCG,GMBCG,AMF,Geach,WHL12 \\
   102   &   356.5763 &     0.9610 &     0.1355 &     0.1324 &  16.06 $\pm$ 0.12 &  13.89 $\pm$ 0.03 &  32.48 &   9.67 &  Geach,WHL12 \\
   103   &   356.8651 &    -0.1538 &     0.2681 &     0.2639 &  16.63 $\pm$ 0.16 &  14.02 $\pm$ 0.03 &  35.66 &   4.08 &  CE,GMBCG,AMF,Geach,WHL12 \\
   104   &   358.7088 &    -0.1407 &     0.1808 &     0.1966 &  15.82 $\pm$ 0.08 &  13.82 $\pm$ 0.03 &  36.48 &   7.75 &  WHL12 \\
   105   &   359.8242 &     0.3167 &     0.1181 &     0.1101 &  16.01 $\pm$ 0.10 &  13.72 $\pm$ 0.03 &  31.60 &  10.75  & New \\ 
   106   &     0.1008 &    -1.2457 &     0.1562 &     0.1618 &  15.26 $\pm$ 0.07 &  13.21 $\pm$ 0.03 &  32.69 &   8.33  & New \\ 
   107   &     0.2157 &    -1.1984 &     0.2170 &     0.1972 &  15.66 $\pm$ 0.09 &  13.68 $\pm$ 0.03 &  30.55 &   5.00  & New \\ 
   108   &     0.3598 &    -0.0288 &     0.2669 &     0.2479 &  16.21 $\pm$ 0.12 &  13.69 $\pm$ 0.03 &  49.82 &   5.75 &  CE,MaxBCG,GMBCG,AMF,Geach,WHL12 \\
   109   &     1.1300 &    -1.1274 &     0.1768 &     0.1788 &  15.82 $\pm$ 0.08 &  13.96 $\pm$ 0.03 &  35.78 &   7.83  & New \\ 
   110   &     1.8501 &    -0.7548 &     0.2325 &     0.1866 &  15.94 $\pm$ 0.09 &  14.02 $\pm$ 0.03 &  51.15 &   7.50 &  CE,Geach \\
   111   &     1.9482 &    -0.7442 &     0.1522 &     0.1070 &  15.67 $\pm$ 0.08 &  13.44 $\pm$ 0.03 &  34.92 &   9.17 &  WHL12 \\
   112   &     2.1160 &    -0.0048 &     0.1216 &     0.1585 &  15.45 $\pm$ 0.07 &  13.38 $\pm$ 0.03 &  33.75 &  11.17  & New \\ 
   113   &     2.4194 &     0.4251 &     0.1904 &     0.1878 &  15.74 $\pm$ 0.09 &  13.64 $\pm$ 0.03 &  35.32 &   7.00 &  CE,Geach,WHL12 \\
   114   &     3.0125 &    -1.0070 &     0.1459 &     0.0848 &  15.71 $\pm$ 0.10 &  13.62 $\pm$ 0.03 &  33.60 &   9.25  & New \\ 
   115   &     3.1982 &     0.7877 &     0.1581 &     0.1484 &  15.60 $\pm$ 0.10 &  13.75 $\pm$ 0.04 &  40.10 &  10.08 &  MaxBCG,WHL12 \\
   116   &     3.2122 &     0.2891 &     0.1430 &     0.1509 &  16.03 $\pm$ 0.10 &  14.17 $\pm$ 0.04 &  47.98 &  13.50 &  CE,MaxBCG,GMBCG,AMF,WHL12 \\
   117   &     3.3711 &     0.6735 &     0.1006 &     0.0843 &  14.98 $\pm$ 0.05 &  13.38 $\pm$ 0.02 &  34.88 &  13.58 &  CE \\
   118   &     4.1540 &    -0.5023 &     0.1393 &     0.1402 &  15.40 $\pm$ 0.08 &  13.76 $\pm$ 0.03 &  44.66 &  12.92 &  CE \\
   119   &     4.2262 &    -1.0625 &     0.2098 &     0.1943 &  15.43 $\pm$ 0.08 &  13.49 $\pm$ 0.03 &  30.02 &   5.17 &  CE,MaxBCG,GMBCG,AMF,Geach,WHL12 \\
   120   &     4.2529 &    -1.2286 &     0.1536 &     0.1617 &  15.54 $\pm$ 0.08 &  13.61 $\pm$ 0.03 &  47.79 &  12.42 &  GMBCG,AMF,WHL12 \\
   121   &     4.4067 &    -0.8784 &     0.2306 &     0.2124 &  16.42 $\pm$ 0.15 &  14.23 $\pm$ 0.03 &  95.39 &  14.17 &  CE,MaxBCG,GMBCG,AMF,Geach,WHL12 \\
   122   &     4.6355 &    -0.7675 &     0.2102 &     0.1924 &  15.85 $\pm$ 0.09 &  13.75 $\pm$ 0.03 &  48.02 &   8.25 &  CE,MaxBCG,GMBCG,Geach,WHL12 \\
   123   &     5.0673 &     0.0794 &     0.2409 &     0.2124 &  15.21 $\pm$ 0.06 &  13.62 $\pm$ 0.06 & 154.31 &  21.33 &  MaxBCG,GMBCG,AMF,Geach,WHL12 \\
   124   &     5.0890 &    -0.2531 &     0.2436 &     0.2100 &  15.75 $\pm$ 0.09 &  13.71 $\pm$ 0.03 &  70.66 &   9.58 &  MaxBCG,AMF,Geach,WHL12 \\
   125   &     5.1726 &    -1.1764 &     0.1989 &     0.1950 &  15.83 $\pm$ 0.10 &  13.80 $\pm$ 0.03 &  37.14 &   6.92 &  MaxBCG,Geach,WHL12 \\
   126   &     5.2040 &     0.1822 &     0.2405 &     0.2139 &  16.31 $\pm$ 0.13 &  13.80 $\pm$ 0.03 & 107.68 &  14.92 &  MaxBCG,GMBCG,WHL12 \\
   127   &     5.3191 &    -0.8368 &     0.1115 &     0.1075 &  15.40 $\pm$ 0.08 &  13.61 $\pm$ 0.03 & 117.18 &  41.92 &  Abell,CE,MaxBCG,GMBCG,Geach,WHL12 \\
   128   &     5.3476 &    -0.8259 &     0.1797 &     0.1675 &  15.40 $\pm$ 0.09 &  13.39 $\pm$ 0.03 & 121.72 &  26.08 &  CE,MaxBCG,GMBCG,AMF,Geach \\
   129   &     5.6779 &    -0.6892 &     0.1867 &     0.1625 &  15.74 $\pm$ 0.07 &  13.99 $\pm$ 0.03 &  37.24 &   7.58 &  MaxBCG,Geach,WHL12 \\
   130   &     5.7616 &    -0.1367 &     0.1726 &     0.1538 &  15.90 $\pm$ 0.11 &  13.91 $\pm$ 0.03 & 107.41 &  24.25 &  Abell,CE,MaxBCG,AMF,Geach,WHL12 \\
        \hline                                                                           
        \hline                                                                          
    \end{tabular}                                                                       
    \small                                                                              
\end{table}

\addtocounter{table}{-1}                                                                
                                                                                        
\begin{table}[htbp]                                                                    
  \centering                                                                             
\footnotesize                                                                          
    \caption{continued}                                                                  
    \begin{tabular}{ c c c c c c c c c c}                                               
      \hline                                                                            
      \hline                                                                            
   ID &    R.A.      &   Decl.     & $z_p$     &  $z_s$      &  $J$    &  $W1$   & $R_{scal}$ & $R_{mea}$  &   Note   \\   
      &   (deg)      &   (deg)     &           &             &  (mag)  & (mag)   &          &            & \\  
  (1) &    (2)       &   (3)       &   (4)     &  (5)        &    (6)  &    (7)	 &    (8)   & (9)        &  (10)      \\      
      \hline                    
   131   &     6.1235 &    -1.0774 &     0.1311 &     0.1387 &  15.99 $\pm$ 0.12 &  14.34 $\pm$ 0.05 &  30.58 &   9.42 &  Geach \\
   132   &     6.3241 &    -0.7247 &     0.1511 &     0.1627 &  15.78 $\pm$ 0.11 &  13.65 $\pm$ 0.03 &  30.51 &   8.08 &  CE,MaxBCG,GMBCG,Geach,WHL12 \\
   133   &     6.6855 &     1.2357 &     0.1673 &     0.1518 &  15.58 $\pm$ 0.10 &  13.52 $\pm$ 0.03 &  49.33 &  11.58 &  Geach,WHL12 \\
   134   &     7.1900 &    -0.0595 &     0.1761 &     0.2131 &  16.01 $\pm$ 0.12 &  14.29 $\pm$ 0.05 &  52.29 &  11.50 &  CE,MaxBCG,GMBCG,AMF,Geach,WHL12 \\
   135   &     7.2203 &    -0.2431 &     0.1504 &     0.1115 &  15.98 $\pm$ 0.12 &  13.48 $\pm$ 0.03 &  57.92 &  15.42 &  CE \\
   136   &     7.2757 &     0.8811 &     0.1131 &     0.1255 &  15.36 $\pm$ 0.09 &  13.39 $\pm$ 0.03 &  30.65 &  10.83  & New \\ 
   137   &     7.4465 &     0.4922 &     0.2091 &     0.1921 &  15.91 $\pm$ 0.09 &  13.80 $\pm$ 0.03 &  48.12 &   8.33 &  WHL12 \\
   138   &     8.0469 &    -0.6669 &     0.2201 &     0.2148 &  16.08 $\pm$ 0.10 &  13.85 $\pm$ 0.03 &  79.63 &  12.75 &  CE,MaxBCG,AMF,Geach,WHL12 \\
   139   &     8.5832 &     0.8158 &     0.2035 &     0.1893 &  16.23 $\pm$ 0.12 &  13.64 $\pm$ 0.03 &  96.65 &  17.42 &  CE,MaxBCG,GMBCG,AMF,Geach,WHL12 \\
   140   &     8.7574 &    -1.2049 &     0.1897 &     0.2118 &  15.94 $\pm$ 0.10 &  13.71 $\pm$ 0.03 &  59.40 &  11.83 &  CE,Geach,WHL12 \\
   141   &     8.7992 &     0.7300 &     0.2688 &     0.2620 &  16.47 $\pm$ 0.14 &  13.98 $\pm$ 0.03 &  49.00 &   5.58 &  CE,GMBCG,Geach,WHL12 \\
   142   &     8.8420 &     1.0672 &     0.2101 &     0.1914 &  15.83 $\pm$ 0.09 &  13.78 $\pm$ 0.03 &  31.98 &   5.50 &  Geach,WHL12 \\
   143   &     9.2985 &     0.0967 &     0.2260 &     0.2555 &  16.44 $\pm$ 0.19 &  13.71 $\pm$ 0.03 &  83.61 &  12.83 &  Geach,WHL12 \\
   144   &     9.6975 &    -1.1918 &     0.1954 &     0.2067 &  15.93 $\pm$ 0.08 &  14.00 $\pm$ 0.03 &  49.31 &   9.42 &  CE,WHL12 \\
   145   &     9.7217 &    -0.2841 &     0.1858 &     0.1820 &  15.86 $\pm$ 0.10 &  13.65 $\pm$ 0.03 &  47.97 &   9.83 &  CE \\
   146   &     9.9280 &     0.6277 &     0.1926 &     0.1462 &  15.76 $\pm$ 0.08 &  13.84 $\pm$ 0.03 &  30.35 &   5.92  & New \\ 
   147   &     9.9306 &    -0.9176 &     0.1072 &     0.1076 &  15.95 $\pm$ 0.10 &  13.80 $\pm$ 0.03 &  43.49 &  16.08 &  CE,WHL12 \\
   148   &    10.8143 &     0.2296 &     0.2117 &     0.2152 &  15.97 $\pm$ 0.11 &  14.18 $\pm$ 0.03 &  41.65 &   7.08 &  MaxBCG,GMBCG \\
   149   &    10.8270 &    -0.3129 &     0.1545 &     0.1525 &  15.15 $\pm$ 0.08 &  13.02 $\pm$ 0.02 &  37.77 &   9.75 &  CE,Geach \\
   150   &    10.8951 &     0.1760 &     0.1693 &     0.1519 &  15.43 $\pm$ 0.08 &  13.57 $\pm$ 0.03 &  86.85 &  20.08 &  Geach \\
   151   &    10.8962 &     1.0196 &     0.2164 &     0.1957 &  15.91 $\pm$ 0.14 &  13.63 $\pm$ 0.03 &  56.77 &   9.33 &  CE,GMBCG,AMF,Geach,WHL12 \\
   152   &    11.0052 &     0.1145 &     0.1869 &     0.2171 &  16.19 $\pm$ 0.14 &  13.78 $\pm$ 0.04 &  81.20 &  16.50 &  CE,MaxBCG,GMBCG,AMF,Geach,WHL12 \\
   153   &    11.0195 &     1.0314 &     0.1160 &     0.1117 &  15.43 $\pm$ 0.09 &  12.91 $\pm$ 0.02 &  43.90 &  15.17 &  Geach,WHL12 \\
   154   &    11.0375 &     1.1917 &     0.1478 &     0.1191 &  15.76 $\pm$ 0.11 &  13.73 $\pm$ 0.04 &  46.37 &  12.58 &  WHL12 \\
   155   &    11.1551 &    -0.9223 &     0.2242 &     0.2006 &  15.95 $\pm$ 0.12 &  13.80 $\pm$ 0.03 &  39.69 &   6.17 &  CE,MaxBCG,GMBCG,Geach,WHL12 \\
   156   &    11.3335 &     0.7793 &     0.1466 &     0.1109 &  16.13 $\pm$ 0.13 &  14.20 $\pm$ 0.05 &  34.97 &   9.58  & New \\ 
   157   &    11.3770 &    -0.7964 &     0.1684 &     0.1472 &  15.10 $\pm$ 0.08 &  13.19 $\pm$ 0.03 &  47.26 &  11.00 &  Abell,CE,MaxBCG,GMBCG,Geach,WHL12 \\
   158   &    11.5601 &     0.0004 &     0.1399 &     0.1117 &  15.88 $\pm$ 0.11 &  14.00 $\pm$ 0.03 &  97.81 &  28.17 &  CE,MaxBCG,GMBCG,AMF,Geach,WHL12 \\
   159   &    11.5934 &    -0.1551 &     0.2373 &     0.2185 &  16.15 $\pm$ 0.13 &  13.67 $\pm$ 0.03 &  61.17 &   8.67 &  CE,MaxBCG,Geach,WHL12 \\
   160   &    11.8588 &    -0.9482 &     0.1523 &     0.1768 &  15.86 $\pm$ 0.08 &  14.02 $\pm$ 0.03 &  79.39 &  20.83 &  Abell,CE \\
   161   &    12.9848 &    -1.0960 &     0.1290 &     0.1342 &  15.20 $\pm$ 0.09 &  13.24 $\pm$ 0.03 &  35.70 &  11.17 &  MaxBCG,AMF,WHL12 \\
   162   &    13.4089 &     0.9207 &     0.2534 &     0.2851 &  16.17 $\pm$ 0.14 &  13.72 $\pm$ 0.03 &  61.19 &   7.75 &  CE,MaxBCG,AMF,Geach,WHL12 \\
   163   &    13.4427 &    -0.7802 &     0.1241 &     0.1376 &  15.39 $\pm$ 0.10 &  13.17 $\pm$ 0.03 &  64.17 &  20.83 &  Abell,CE,GMBCG,Geach,WHL12 \\
   164   &    13.8141 &    -0.3482 &     0.1635 &     0.1462 &  15.44 $\pm$ 0.07 &  13.65 $\pm$ 0.03 &  51.77 &  12.50 &  CE,MaxBCG,GMBCG,Geach,WHL12 \\
   165   &    14.3250 &     0.0257 &     0.1912 &     0.1935 &  15.75 $\pm$ 0.11 &  13.59 $\pm$ 0.03 &  54.97 &  10.83 &  CE,Geach,WHL12 \\
   166   &    15.2043 &    -0.4221 &     0.2696 &     0.2022 &  15.86 $\pm$ 0.09 &  13.81 $\pm$ 0.03 &  60.31 &   6.83  & New \\ 
   167   &    15.2281 &    -1.1552 &     0.1412 &      ...   &  15.00 $\pm$ 9.99 &  13.90 $\pm$ 0.03 &  47.60 &  13.58  & New \\ 
   168   &    15.3084 &     0.5743 &     0.2113 &     0.1994 &  15.63 $\pm$ 0.09 &  13.52 $\pm$ 0.03 &  70.42 &  12.00 &  CE,MaxBCG,GMBCG,AMF,Geach,WHL12 \\
   169   &    15.3301 &    -0.0553 &     0.1856 &     0.1935 &  15.47 $\pm$ 0.09 &  13.76 $\pm$ 0.05 &  56.85 &  11.67 &  CE \\
   170   &    15.3609 &    -0.0636 &     0.1136 &     0.1078 &  14.90 $\pm$ 0.08 &  12.80 $\pm$ 0.02 &  48.06 &  16.92 &  Abell,CE,Geach,WHL12 \\   
   171   &    15.3715 &    -0.2619 &     0.2234 &     0.1929 &  15.71 $\pm$ 0.09 &  13.75 $\pm$ 0.03 &  36.26 &   5.67 &  WHL12 \\
   172   &    15.3876 &     0.5376 &     0.1114 &     0.1176 &  15.38 $\pm$ 0.09 &  13.61 $\pm$ 0.03 &  35.13 &  12.58 &  WHL12 \\
   173   &    15.4146 &    -0.2248 &     0.1290 &     0.1111 &  15.56 $\pm$ 0.07 &  13.78 $\pm$ 0.03 &  50.59 &  15.83 &  Abell,CE,WHL12 \\
   174   &    15.4679 &     0.6898 &     0.1292 &     0.1456 &  15.94 $\pm$ 0.09 &  13.98 $\pm$ 0.03 &  30.16 &   9.42 &  CE,WHL12 \\
   175   &    15.6797 &     1.1362 &     0.1546 &     0.1440 &  15.41 $\pm$ 0.10 &  13.03 $\pm$ 0.02 &  65.92 &  17.00 &  CE,MaxBCG,Geach,WHL12 \\
   176   &    15.7192 &     0.2479 &     0.2094 &     0.2047 &  15.79 $\pm$ 0.10 &  13.87 $\pm$ 0.03 &  61.22 &  10.58 &  CE,Geach,WHL12 \\
   177   &    16.0185 &    -0.4345 &     0.2879 &     0.2790 &  16.04 $\pm$ 0.10 &  13.79 $\pm$ 0.03 &  46.72 &   4.67 &  CE,MaxBCG,GMBCG,Geach,WHL12 \\
   178   &    16.2306 &     0.0602 &     0.2903 &     0.2721 &  16.24 $\pm$ 0.12 &  13.79 $\pm$ 0.03 &  86.44 &   8.50 &  CE,MaxBCG,GMBCG,Geach,WHL12 \\
   179   &    16.6342 &     0.6310 &     0.1578 &     0.1447 &  16.13 $\pm$ 0.12 &  14.01 $\pm$ 0.03 &  38.73 &   9.75 &  WHL12 \\
   180   &    16.8165 &     0.6614 &     0.1496 &     0.1541 &  15.55 $\pm$ 0.08 &  13.64 $\pm$ 0.03 &  37.97 &  10.17 &  CE \\
   181   &    16.8618 &     0.1470 &     0.2597 &     0.2515 &  16.11 $\pm$ 0.11 &  13.99 $\pm$ 0.03 &  46.03 &   5.58 &  CE,MaxBCG,AMF,Geach,WHL12 \\
   182   &    16.9189 &     1.0481 &     0.1483 &     0.1553 &  15.52 $\pm$ 0.10 &  13.49 $\pm$ 0.03 &  31.44 &   8.50 &  WHL12 \\
   183   &    17.3901 &    -0.8986 &     0.2063 &     0.1737 &  15.45 $\pm$ 0.08 &  13.51 $\pm$ 0.03 &  59.87 &  10.58 &  CE,Geach,WHL12 \\
   184   &    17.4051 &    -0.9297 &     0.1210 &     0.0882 &  16.16 $\pm$ 0.12 &  13.91 $\pm$ 0.04 &  59.65 &  19.83  & New \\ 
   185   &    17.6656 &     1.0666 &     0.1926 &     0.1764 &  15.44 $\pm$ 0.08 &  13.61 $\pm$ 0.03 &  44.86 &   8.75 &  CE,MaxBCG,AMF,WHL12 \\
   186   &    17.9198 &    -0.7625 &     0.1116 &     0.1318 &  15.79 $\pm$ 0.11 &  13.90 $\pm$ 0.03 &  43.14 &  15.42 &  Geach,WHL12 \\
   187   &    17.9535 &    -0.0181 &     0.2379 &     0.2538 &  15.89 $\pm$ 0.12 &  13.46 $\pm$ 0.03 &  97.41 &  13.75 &  CE,MaxBCG,GMBCG,Geach,WHL12 \\
   188   &    18.1542 &    -0.6651 &     0.2357 &     0.2467 &  15.85 $\pm$ 0.09 &  13.79 $\pm$ 0.03 &  52.88 &   7.58 &  MaxBCG,Geach \\
   189   &    18.2290 &     1.0638 &     0.1416 &     0.1333 &  15.62 $\pm$ 0.07 &  13.81 $\pm$ 0.03 &  34.01 &   9.67  & New \\ 
   190   &    18.2797 &    -0.0010 &     0.2128 &     0.2156 &  15.99 $\pm$ 0.10 &  13.84 $\pm$ 0.03 &  30.13 &   5.08 &  Geach,WHL12 \\
   191   &    18.5621 &    -0.9125 &     0.2074 &     0.1835 &  15.56 $\pm$ 0.08 &  13.58 $\pm$ 0.03 &  55.63 &   9.75 &  CE,MaxBCG,GMBCG,Geach,WHL12 \\
   192   &    18.6568 &    -0.8458 &     0.2023 &     0.1820 &  15.94 $\pm$ 0.09 &  13.98 $\pm$ 0.03 &  61.46 &  11.17 &  GMBCG,AMF,WHL12 \\
   193   &    18.8958 &    -0.4914 &     0.1554 &     0.1828 &  15.98 $\pm$ 0.11 &  13.87 $\pm$ 0.04 &  35.77 &   9.17 &  Geach,WHL12 \\
   194   &    19.1019 &    -0.1334 &     0.1943 &     0.1766 &  15.45 $\pm$ 0.09 &  13.39 $\pm$ 0.03 &  45.84 &   8.83 &  CE,MaxBCG,Geach,WHL12 \\
   195   &    19.4654 &    -1.2269 &     0.2377 &     0.2171 &  15.97 $\pm$ 0.10 &  13.99 $\pm$ 0.03 &  37.13 &   5.25 &  Geach,WHL12 \\
        \hline                                                                           
        \hline                                                                          
    \end{tabular}                                                                       
    \small                                                                              
\end{table}

\addtocounter{table}{-1}                                                                
                                                                                        
\begin{table}[htbp]                                                                    
  \centering                                                                             
\footnotesize                                                                          
    \caption{continued}                                                                  
    \begin{tabular}{ c c c c c c c c c c}                                               
      \hline                                                                            
      \hline                                                                            
   ID &    R.A.      &   Decl.     & $z_p$     &  $z_s$      &  $J$    &  $W1$   & $R_{scal}$ & $R_{mea}$  &   Note   \\   
      &   (deg)      &   (deg)     &           &             &  (mag)  & (mag)   &          &            & \\  
  (1) &    (2)       &   (3)       &   (4)     &  (5)        &    (6)  &    (7)	 &    (8)   & (9)        &  (10)      \\      
      \hline                    
   196   &    19.7274 &    -1.0984 &     0.1276 &     0.1216 &  15.02 $\pm$ 0.07 &  13.18 $\pm$ 0.03 &  35.83 &  11.33  & New \\ 
   197   &    19.7734 &    -1.2336 &     0.1260 &     0.1220 &  15.21 $\pm$ 0.08 &  13.45 $\pm$ 0.05 &  55.45 &  17.75  & New \\ 
   198   &    19.8598 &    -0.7436 &     0.2421 &     0.2172 &  15.68 $\pm$ 0.08 &  13.66 $\pm$ 0.03 &  39.56 &   5.42 &  CE,MaxBCG,Geach,WHL12 \\
   199   &    19.8798 &    -1.1572 &     0.2036 &     0.1860 &  16.07 $\pm$ 0.11 &  14.10 $\pm$ 0.03 &  74.52 &  13.42 &  CE,Geach,WHL12 \\
   200   &    20.1750 &    -0.6050 &     0.1124 &     0.0938 &  15.88 $\pm$ 0.10 &  13.84 $\pm$ 0.03 &  37.53 &  13.33 &  WHL12 \\
   201   &    20.4632 &    -0.1918 &     0.1979 &     0.2001 &  15.69 $\pm$ 0.10 &  13.58 $\pm$ 0.03 &  31.96 &   6.00 &  CE,AMF,Geach,WHL12 \\
   202   &    20.4820 &     0.0662 &     0.1347 &     0.0774 &  15.24 $\pm$ 0.09 &  13.20 $\pm$ 0.03 &  32.29 &   9.67  & New \\ 
   203   &    20.5108 &     0.3345 &     0.1551 &     0.1745 &  15.53 $\pm$ 0.11 &  13.17 $\pm$ 0.03 &  80.74 &  20.75 &  Abell,CE,MaxBCG,GMBCG,AMF,Geach,WHL12 \\
   204   &    20.6525 &    -0.8140 &     0.1797 &     0.1726 &  15.65 $\pm$ 0.10 &  13.38 $\pm$ 0.03 &  63.38 &  13.58 &  CE,MaxBCG,GMBCG,AMF,Geach,WHL12 \\
   205   &    20.8281 &     1.1489 &     0.1298 &     0.1076 &  15.80 $\pm$ 0.07 &  14.07 $\pm$ 0.03 &  35.12 &  10.92  & New \\ 
   206   &    21.0764 &    -1.2405 &     0.1728 &     0.1725 &  15.56 $\pm$ 0.08 &  13.43 $\pm$ 0.02 &  51.02 &  11.50 &  GMBCG,Geach,WHL12 \\
   207   &    21.2476 &    -0.8441 &     0.1627 &     0.1707 &  15.68 $\pm$ 0.08 &  13.95 $\pm$ 0.03 &  32.61 &   7.92  & New \\ 
   208   &    21.5269 &     1.2269 &     0.1925 &     0.2093 &  15.48 $\pm$ 9.99 &  14.56 $\pm$ 0.05 &  35.45 &   6.92 &  CE,AMF,Geach,WHL12 \\
   209   &    21.7496 &    -0.7839 &     0.1807 &     0.1647 &  15.66 $\pm$ 0.08 &  13.76 $\pm$ 0.03 &  31.37 &   6.67  & New \\ 
   210   &    23.2450 &     0.9407 &     0.1250 &     0.1226 &  15.38 $\pm$ 0.09 &  13.48 $\pm$ 0.03 &  42.62 &  13.75 &  CE,MaxBCG,Geach,WHL12 \\
   211   &    23.7048 &    -0.6121 &     0.1557 &     0.0838 &  15.10 $\pm$ 0.06 &  12.99 $\pm$ 0.03 &  38.10 &   9.75 &  CE,GMBCG,AMF,Geach,WHL12 \\
   212   &    23.7313 &     0.3878 &     0.1592 &     0.1535 &  15.74 $\pm$ 0.11 &  13.66 $\pm$ 0.03 &  40.13 &  10.00 &  MaxBCG,AMF,Geach,WHL12 \\
   213   &    23.7849 &    -1.1513 &     0.1649 &     0.1555 &  15.73 $\pm$ 0.08 &  13.78 $\pm$ 0.03 &  44.98 &  10.75 &  CE,GMBCG,Geach,WHL12 \\
   214   &    23.8864 &     0.3775 &     0.1486 &     0.1523 &  15.66 $\pm$ 0.10 &  13.77 $\pm$ 0.03 &  34.93 &   9.42 &  Geach \\
   215   &    23.9820 &     0.5584 &     0.1160 &     0.1351 &  15.33 $\pm$ 0.07 &  13.58 $\pm$ 0.03 &  33.28 &  11.50  & New \\ 
   216   &    24.3334 &     0.9561 &     0.1780 &     0.1976 &  15.85 $\pm$ 0.09 &  13.74 $\pm$ 0.03 &  46.46 &  10.08  & New \\ 
   217   &    24.7484 &    -0.8523 &     0.1202 &     0.1179 &  15.53 $\pm$ 0.09 &  13.61 $\pm$ 0.03 &  44.34 &  14.83  & New \\ 
   218   &    24.8218 &     0.3343 &     0.2212 &     0.1966 &  15.97 $\pm$ 0.12 &  13.89 $\pm$ 0.03 &  36.21 &   5.75  & New \\ 
   219   &    25.2334 &     0.1560 &     0.1758 &     0.1677 &  15.58 $\pm$ 0.09 &  13.54 $\pm$ 0.03 &  51.02 &  11.25 &  CE,Geach,WHL12 \\
   220   &    25.3794 &    -0.9283 &     0.1596 &     0.1547 &  15.60 $\pm$ 0.08 &  13.86 $\pm$ 0.03 &  36.89 &   9.17 &  CE,Geach \\
   221   &    25.4900 &    -1.1074 &     0.1653 &     0.1560 &  15.20 $\pm$ 0.09 &  13.15 $\pm$ 0.02 &  46.18 &  11.00 &  CE,MaxBCG,AMF,Geach,WHL12 \\
   222   &    25.6666 &     0.1151 &     0.1120 &     0.1010 &  15.44 $\pm$ 0.06 &  13.77 $\pm$ 0.03 &  33.44 &  11.92  & New \\ 
   223   &    25.6991 &     0.8671 &     0.1118 &     0.1015 &  15.61 $\pm$ 0.09 &  13.52 $\pm$ 0.03 &  31.76 &  11.33  & New \\ 
   224   &    25.7110 &     0.7474 &     0.2027 &     0.1962 &  15.81 $\pm$ 0.10 &  13.91 $\pm$ 0.03 &  35.41 &   6.42 &  CE,WHL12 \\
   225   &    26.2515 &    -0.8150 &     0.2335 &     0.1968 &  15.95 $\pm$ 0.10 &  13.93 $\pm$ 0.03 &  50.36 &   7.33 &  CE,WHL12 \\
   226   &    26.3010 &    -0.0931 &     0.2044 &     0.1989 &  15.48 $\pm$ 0.08 &  13.41 $\pm$ 0.03 &  34.91 &   6.25 &  CE,GMBCG,Geach,WHL12 \\
   227   &    26.6881 &    -0.6752 &     0.1002 &     0.0823 &  15.51 $\pm$ 0.08 &  13.65 $\pm$ 0.03 &  44.83 &  17.50 &  AMF,WHL12 \\
   228   &    27.0689 &     0.3583 &     0.1826 &     0.2061 &  ...              &  13.92 $\pm$ 0.03 &  50.05 &  10.50 &  CE \\
   229   &    27.7459 &    -1.0235 &     0.1752 &     0.1583 &  15.93 $\pm$ 0.10 &  13.85 $\pm$ 0.03 &  36.90 &   8.17 &  Geach \\
   230   &    27.8014 &    -0.9955 &     0.2466 &     0.2428 &  16.07 $\pm$ 0.09 &  13.74 $\pm$ 0.03 &  57.76 &   7.67 &  CE,WHL12 \\
   231   &    28.1525 &    -0.2279 &     0.1746 &     0.1768 &  15.39 $\pm$ 0.08 &  13.43 $\pm$ 0.02 &  32.96 &   7.33  & New \\ 
   232   &    28.1750 &     1.0072 &     0.2132 &     0.2297 &  16.07 $\pm$ 0.15 &  13.82 $\pm$ 0.04 &  67.37 &  11.33 &  Abell,CE,MaxBCG,GMBCG,AMF,Geach,WHL12 \\
   233   &    28.1897 &    -0.3932 &     0.1534 &     0.1770 &  15.68 $\pm$ 0.09 &  13.81 $\pm$ 0.03 &  43.52 &  11.33  & New \\ 
   234   &    28.2245 &    -0.8960 &     0.1141 &     0.1192 &  15.17 $\pm$ 0.09 &  13.27 $\pm$ 0.03 &  50.87 &  17.83  & New \\ 
   235   &    28.2534 &    -0.5705 &     0.1317 &     0.1331 &  16.04 $\pm$ 0.12 &  13.87 $\pm$ 0.03 &  71.54 &  21.92 &  CE,MaxBCG,WHL12 \\        
   236   &    28.3570 &    -1.1600 &     0.2138 &     0.2416 &  15.85 $\pm$ 0.10 &  13.59 $\pm$ 0.03 & 158.35 &  26.50 &  Abell,CE,MaxBCG,GMBCG,AMF,Geach,WHL12 \\
   237   &    29.1181 &     1.0605 &     0.1719 &     0.1826 &  15.64 $\pm$ 0.08 &  13.70 $\pm$ 0.03 & 140.27 &  31.83  & New \\ 
   238   &    29.3485 &     0.4178 &     0.1462 &     0.1350 &  15.07 $\pm$ 0.07 &  13.18 $\pm$ 0.02 &  32.15 &   8.83 &  CE,MaxBCG,GMBCG,Geach,WHL12 \\
   239   &    29.4308 &    -0.1504 &     0.1246 &     0.1349 &  15.85 $\pm$ 0.10 &  13.95 $\pm$ 0.04 &  38.89 &  12.58 &  MaxBCG,GMBCG,Geach,WHL12 \\
   240   &    29.4772 &    -0.6338 &     0.1914 &     0.1884 &  15.97 $\pm$ 0.08 &  13.95 $\pm$ 0.03 &  91.52 &  18.00 &  Abell,CE,GMBCG,AMF,Geach,WHL12 \\
   241   &    29.4970 &    -0.7244 &     0.2162 &     0.1865 &  15.78 $\pm$ 0.09 &  13.72 $\pm$ 0.03 &  86.56 &  14.25 &  Abell,CE,MaxBCG,GMBCG,AMF,Geach,WHL12 \\
   242   &    29.7756 &     0.8355 &     0.1443 &     0.1354 &  15.58 $\pm$ 0.08 &  13.73 $\pm$ 0.03 &  32.01 &   8.92 &  Geach,WHL12 \\
   243   &    29.8175 &    -0.1096 &     0.1626 &     0.1550 &  15.61 $\pm$ 0.07 &  13.84 $\pm$ 0.03 &  46.98 &  11.42 &  CE,MaxBCG,Geach,WHL12 \\
   244   &    30.0376 &    -0.8842 &     0.2128 &     0.2089 &  16.18 $\pm$ 0.10 &  13.69 $\pm$ 0.03 &  38.55 &   6.50 &  WHL12 \\
   245   &    30.3347 &    -0.4569 &     0.1619 &     0.1597 &  15.45 $\pm$ 0.08 &  13.47 $\pm$ 0.02 &  33.07 &   8.08 &  CE,MaxBCG,GMBCG,AMF,Geach,WHL12 \\
   246   &    30.5012 &    -0.4818 &     0.1460 &     0.1589 &  15.71 $\pm$ 0.09 &  13.78 $\pm$ 0.03 &  40.60 &  11.17 &  Geach \\
   247   &    31.0914 &    -0.8329 &     0.1526 &     0.1367 &  15.54 $\pm$ 0.08 &  13.63 $\pm$ 0.03 &  30.24 &   7.92  & New \\ 
   248   &    31.1257 &     0.3046 &     0.2324 &     0.1726 &  16.25 $\pm$ 0.15 &  13.75 $\pm$ 0.03 &  64.19 &   9.42 &  Abell,CE,MaxBCG,AMF,Geach,WHL12 \\
   249   &    31.1304 &     0.2282 &     0.1490 &     0.1631 &  16.26 $\pm$ 0.15 &  13.88 $\pm$ 0.04 &  33.76 &   9.08 &  CE,MaxBCG,Geach,WHL12 \\
   250   &    31.1345 &    -1.0540 &     0.1699 &     0.1617 &  16.23 $\pm$ 0.14 &  13.82 $\pm$ 0.03 &  43.44 &  10.00 &  CE,Geach \\
   251   &    31.3167 &     0.0574 &     0.1331 &     0.1132 &  15.75 $\pm$ 0.09 &  13.88 $\pm$ 0.03 &  49.47 &  15.00  & New \\ 
   252   &    31.3794 &     0.1897 &     0.1826 &     0.1716 &  15.37 $\pm$ 0.09 &  13.35 $\pm$ 0.02 &  30.17 &   6.33 &  CE,MaxBCG,AMF,Geach,WHL12 \\
   253   &    31.4742 &     0.0331 &     0.1821 &     0.1735 &  15.42 $\pm$ 0.07 &  13.30 $\pm$ 0.03 &  63.33 &  13.33 &  Geach,WHL12 \\
   254   &    32.4464 &    -0.1118 &     0.1639 &     0.1521 &  15.54 $\pm$ 0.09 &  13.52 $\pm$ 0.03 &  36.70 &   8.83 &  CE,Geach,WHL12 \\
   255   &    32.5758 &    -1.0184 &     0.1842 &     0.1709 &  15.66 $\pm$ 0.08 &  13.75 $\pm$ 0.03 &  79.58 &  16.50 &  Abell,CE,MaxBCG,AMF,Geach,WHL12 \\
   256   &    32.7266 &    -1.1567 &     0.1432 &     0.1760 &  15.86 $\pm$ 0.15 &  13.68 $\pm$ 0.03 &  52.52 &  14.75 &  CE,Geach,WHL12 \\
   257   &    32.8846 &     0.1167 &     0.2268 &     0.2122 &  16.57 $\pm$ 0.14 &  14.33 $\pm$ 0.03 &  40.96 &   6.25 &  CE,MaxBCG,GMBCG,Geach,WHL12 \\
   258   &    33.0959 &    -0.4210 &     0.1177 &     0.1015 &  15.37 $\pm$ 0.08 &  13.39 $\pm$ 0.02 &  32.49 &  11.08 &  Geach \\
   259   &    33.1749 &     0.4748 &     0.2069 &     0.2015 &  15.91 $\pm$ 0.11 &  13.84 $\pm$ 0.03 &  58.30 &  10.25 &  CE,MaxBCG,AMF,Geach,WHL12 \\
   260   &    33.4638 &     0.4673 &     0.1940 &     0.1820 &  15.79 $\pm$ 0.11 &  13.66 $\pm$ 0.03 &  44.43 &   8.58 &  CE,MaxBCG,GMBCG,AMF,Geach,WHL12 \\
        \hline                                                                           
        \hline                                                                          
    \end{tabular}                                                                       
    \small                                                                              
\end{table}

\addtocounter{table}{-1}                                                                
                                                                                        
\begin{table}[htbp]                                                                    
  \centering                                                                             
\footnotesize                                                                          
    \caption{continued}                                                                  
    \begin{tabular}{ c c c c c c c c c c}                                               
      \hline                                                                            
      \hline                                                                            
   ID &    R.A.      &   Decl.     & $z_p$     &  $z_s$      &  $J$    &  $W1$   & $R_{scal}$ & $R_{mea}$  &   Note   \\   
      &   (deg)      &   (deg)     &           &             &  (mag)  & (mag)   &          &            & \\  
  (1) &    (2)       &   (3)       &   (4)     &  (5)        &    (6)  &    (7)	 &    (8)   & (9)        &  (10)      \\      
      \hline                    
   261   &    33.4849 &     0.5268 &     0.2567 &     0.2126 &  15.91 $\pm$ 0.10 &  13.53 $\pm$ 0.03 &  94.30 &  11.67 &  CE,MaxBCG,GMBCG,AMF,Geach,WHL12 \\
   262   &    33.5527 &    -0.1887 &     0.1381 &     0.1408 &  15.59 $\pm$ 0.11 &  13.69 $\pm$ 0.04 &  57.35 &  16.75 &  CE,MaxBCG,GMBCG,AMF,Geach,WHL12 \\
   263   &    33.7983 &     1.0014 &     0.1095 &     0.1223 &  15.18 $\pm$ 0.08 &  13.23 $\pm$ 0.02 &  43.37 &  15.75 &  CE \\
   264   &    33.7986 &     0.9053 &     0.1719 &     0.1219 &  15.96 $\pm$ 0.09 &  14.05 $\pm$ 0.03 &  74.92 &  17.00  & New \\ 
   265   &    34.6748 &     0.1138 &     0.2509 &     0.2719 &  16.30 $\pm$ 0.16 &  13.66 $\pm$ 0.03 &  71.14 &   9.17 &  CE,MaxBCG,GMBCG,AMF,Geach,WHL12 \\
   266   &    36.3090 &    -1.0862 &     0.1841 &     0.1694 &  15.70 $\pm$ 0.10 &  13.83 $\pm$ 0.03 &  31.32 &   6.50 &  CE,GMBCG,Geach,WHL12 \\
   267   &    36.7314 &    -1.0641 &     0.1227 &     0.0967 &  15.63 $\pm$ 0.10 &  13.70 $\pm$ 0.03 &  47.97 &  15.75  & New \\ 
   268   &    37.3373 &     0.5937 &     0.1108 &     0.1316 &  15.12 $\pm$ 0.08 &  13.20 $\pm$ 0.03 &  51.22 &  18.42 &  WHL12 \\
   269   &    37.7454 &     1.1128 &     0.1787 &     0.1495 &  15.61 $\pm$ 0.09 &  13.52 $\pm$ 0.03 &  49.79 &  10.75 &  CE \\
   270   &    38.4721 &     0.0777 &     0.1856 &     0.1857 &  15.30 $\pm$ 0.09 &  13.43 $\pm$ 0.04 &  42.63 &   8.75 &  CE,Geach,WHL12 \\
   271   &    40.2131 &    -0.9324 &     0.2561 &     0.2396 &  16.31 $\pm$ 0.14 &  13.78 $\pm$ 0.03 &  40.21 &   5.00 &  CE,MaxBCG,AMF,Geach,WHL12 \\
   272   &    40.6842 &    -0.9660 &     0.1255 &     0.1797 &  15.64 $\pm$ 0.09 &  13.36 $\pm$ 0.03 &  31.91 &  10.25 &  MaxBCG \\
   273   &    40.8013 &    -1.0201 &     0.2595 &     0.2386 &  15.86 $\pm$ 0.11 &  13.67 $\pm$ 0.03 &  90.60 &  11.00 &  CE,GMBCG,AMF,Geach,WHL12 \\
   274   &    41.4178 &    -0.7229 &     0.2143 &     0.1821 &  15.65 $\pm$ 0.09 &  13.54 $\pm$ 0.03 &  51.98 &   8.67 &  CE,MaxBCG,GMBCG,AMF,Geach,WHL12 \\
   275   &    41.8485 &     0.1005 &     0.1944 &     0.1810 &  15.33 $\pm$ 0.07 &  13.29 $\pm$ 0.03 &  35.49 &   6.83 &  Geach,WHL12 \\
   276   &    42.3182 &    -0.7739 &     0.1493 &     0.1376 &  15.13 $\pm$ 0.07 &  13.28 $\pm$ 0.02 &  45.02 &  12.08 &  WHL12 \\
   277   &    42.3603 &     0.0250 &     0.1821 &     0.1503 &  15.55 $\pm$ 0.10 &  13.44 $\pm$ 0.02 &  35.63 &   7.50 &  CE,MaxBCG,GMBCG,AMF,Geach,WHL12 \\
   278   &    42.4043 &     0.2698 &     0.1839 &     0.1770 &  15.95 $\pm$ 0.10 &  13.59 $\pm$ 0.03 &  30.91 &   6.42  & New \\ 
   279   &    42.9448 &    -0.1646 &     0.2133 &     0.1832 &  15.62 $\pm$ 0.09 &  13.66 $\pm$ 0.03 &  34.69 &   5.83 &  Geach,WHL12 \\
   280   &    43.1920 &     1.0847 &     0.1325 &     0.1374 &  15.42 $\pm$ 0.09 &  13.27 $\pm$ 0.02 &  81.28 &  24.75 &  CE,Geach,WHL12 \\
   281   &    44.7325 &     0.2605 &     0.1181 &     0.1277 &  15.53 $\pm$ 0.08 &  13.68 $\pm$ 0.03 &  51.21 &  17.42 &  WHL12 \\
   282   &    44.8857 &     0.2318 &     0.1828 &     0.1933 &  15.30 $\pm$ 0.09 &  12.97 $\pm$ 0.02 &  46.93 &   9.83 &  CE,MaxBCG,GMBCG,AMF,Geach,WHL12 \\
   283   &    45.6562 &    -0.6689 &     0.2207 &     0.1900 &  16.23 $\pm$ 0.13 &  14.01 $\pm$ 0.03 &  39.23 &   6.25 &  CE,Geach,WHL12 \\
   284   &    45.7460 &    -0.1935 &     0.1656 &     0.1571 &  15.48 $\pm$ 0.08 &  13.60 $\pm$ 0.03 &  61.70 &  14.67 &  CE,MaxBCG,AMF,Geach,WHL12 \\
   285   &    46.0997 &     0.8304 &     0.1099 &     0.1351 &  15.92 $\pm$ 0.10 &  13.78 $\pm$ 0.03 &  32.91 &  11.92  & New \\ 
   286   &    46.1983 &    -0.9336 &     0.1550 &     0.1626 &  15.72 $\pm$ 0.08 &  13.84 $\pm$ 0.03 &  32.73 &   8.42  & New \\ 
   287   &    46.6213 &     0.9996 &     0.1493 &     0.1536 &  15.59 $\pm$ 0.08 &  13.54 $\pm$ 0.02 &  30.75 &   8.25 &  CE,Geach \\
   288   &    47.2143 &    -1.1733 &     0.1231 &     0.1262 &  15.50 $\pm$ 0.08 &  13.65 $\pm$ 0.02 &  31.57 &  10.33 &  Geach \\
   289   &    48.3834 &     0.9172 &     0.1537 &     0.1425 &  16.16 $\pm$ 0.11 &  13.78 $\pm$ 0.03 &  34.66 &   9.00 &  CE,WHL12 \\
   290   &    48.5910 &    -0.5879 &     0.1228 &     0.1167 &  15.66 $\pm$ 0.08 &  13.85 $\pm$ 0.03 &  67.06 &  22.00 &  MaxBCG,AMF,Geach,WHL12 \\
   291   &    48.6184 &     0.2598 &     0.1473 &     0.1294 &  15.81 $\pm$ 0.09 &  13.85 $\pm$ 0.03 &  46.17 &  12.58 &  CE,Geach,WHL12 \\
   292   &    49.3038 &     0.1131 &     0.1230 &     0.1147 &  15.06 $\pm$ 0.08 &  13.20 $\pm$ 0.03 &  35.88 &  11.75 &  WHL12 \\
   293   &    49.7007 &     0.6802 &     0.2113 &     0.1746 &  16.37 $\pm$ 0.12 &  14.01 $\pm$ 0.03 &  35.21 &   6.00 &  CE,WHL12 \\
   294   &    50.1673 &    -1.2419 &     0.1836 &     0.1812 &  16.03 $\pm$ 0.10 &  13.82 $\pm$ 0.03 &  30.83 &   6.42 &  CE,Geach,WHL12 \\
   295   &    50.5087 &     1.1235 &     0.2324 &     0.1489 &  15.33 $\pm$ 9.99 &  13.71 $\pm$ 0.03 &  32.91 &   4.83 &  CE,Geach \\
   296   &    50.7242 &     0.7819 &     0.1220 &     0.1487 &  15.73 $\pm$ 0.09 &  14.05 $\pm$ 0.04 &  33.84 &  11.17 &  CE,Geach \\
   297   &    52.9716 &    -0.7835 &     0.1547 &     0.1368 &  15.34 $\pm$ 0.07 &  13.34 $\pm$ 0.02 &  43.00 &  11.08 &  CE,GMBCG,AMF,Geach,WHL12 \\
   298   &    53.5507 &     1.1746 &     0.1562 &     0.1642 &  15.63 $\pm$ 0.13 &  13.82 $\pm$ 0.05 &  49.72 &  12.67 &  CE,Geach \\
   299   &    53.6419 &    -1.1646 &     0.1409 &     0.1389 &  15.43 $\pm$ 0.10 &  13.20 $\pm$ 0.03 &  73.76 &  21.08 &  CE,GMBCG,Geach,WHL12 \\
   300   &    54.9725 &    -0.2834 &     0.1392 &     0.1274 &  15.44 $\pm$ 0.11 &  13.04 $\pm$ 0.03 &  43.17 &  12.50 &  WHL12 \\                   
   301   &    55.6778 &    -0.2856 &     0.2795 &     0.3072 &  16.60 $\pm$ 0.16 &  13.46 $\pm$ 0.03 &  58.30 &   6.17 &  CE,MaxBCG,GMBCG,AMF,Geach,WHL12 \\
   302   &    58.2873 &    -0.8459 &     0.1212 &     0.1319 &  15.59 $\pm$ 0.09 &  13.77 $\pm$ 0.03 &  34.11 &  11.33 &  WHL12 \\
       \hline
       \hline
    \end{tabular}
    \small
\\
{
    Note. 
    Column(1): the sequence number; 
    Column(2): R.A. (J2000) of cluster center in degree; 
    Column(3): Decl. (J2000) of cluster center in degree; 
    Column(4): photometric redshift of cluster with a typical uncertainty of 0.022; 
    Column(5): spectroscopic redshift of central galaxy with a typical uncertainty of $2\times10^{-4}$; 
    Column(6): $J$-band magnitude of centeral galaxy; 
    Column(7): $W1$-band magnitude of centeral galaxy; 
    Column(8): scaled richness of cluster; 
    Column(9): measured richness of cluster.
    Column(10): Notes for known clusters in other catalogs: 
Abell (Abell 1958; Abell et al. 1989); 
CE (Goto et al. 2002); 
MaxBCG (Koester et al. 2007); 
GMBCG (Hao et al. 2010); 
AMF (Szabo et al. 2011); 
Geach (Geach et al. 2011), 
WHL12 (Wen et al. 2012); 
newly identified clusters are labelled ``New''.}
\end{table}

\end{document}